\documentclass[fleqn,usenatbib,useAMS]{mnras}
\usepackage{amsmath}	
\usepackage{amssymb}	
\usepackage{multicol}        
\usepackage{mathtools}
\usepackage{bm}		
\usepackage{graphicx,epsfig}
\usepackage{float}
\usepackage{xcolor}

\usepackage[T1]{fontenc}
\usepackage{ae,aecompl}

\title[Outer potential of elliptical torus]{Outer gravitational potential of a homogeneous torus with an elliptical cross-section: I. Representation by two massive circles}
\author[E. Yu. Bannikova, \& S. V. Skolota] {Elena Yu. Bannikova$^{1,2,3}$\thanks{Contact e-mail: \href{mailto:bannikova@astron.kharkov.us}
{bannikova@astron.kharkov.ua}},  Sergey V. Skolota$^{3}$\\
$^{1}$ INAF - Astronomical Observatory of Capodimonte, Salita Moiariello 16, I-80131, Naples, Italy \\
$^{2}$ Institute of Radio Astronomy, National Academy of Sciences of Ukraine, Mystetstv 4, UA-61002 Kharkiv, Ukraine \\
$^{3}$ V.N.Karazin Kharkiv National University, Svobody Sq.4, UA-61022, Kharkiv, Ukraine 
}

\date{Last updated ****  **; in original form ****  *}

\pubyear{2020}

\begin{document}
\label{firstpage}
\pagerange{\pageref{firstpage}--\pageref{lastpage}}
\maketitle

\begin{abstract}
This paper deals with the gravitational potential of a homogeneous torus with elliptical cross-section. We present a new expression for its gravitational potential which is valid in any point of the space, obtained by modeling the torus with a set of massive circles (infinitely thin rings). 
We found that the outer potential can be represented with good accuracy by the potential of two massive circles with masses which are half of the torus mass. These massive circles intercept the elliptical cross-section at two points along the major axis which are in opposite directions and at half of the distances to the foci of the cross-section. The same formula works for both cases: oblate and prolate cross-sections. For the case of the prolate cross-section of the torus the distances to massive circles are imaginary and conjugate ones but the values of the torus potential for this case are real. The obtained approximation is robust as the error maps show. 
\end{abstract}



\section{Introduction}
\label{Intro}
The beginning of investigation of the torus potential is related with the paper of B.Riemann \citep{Riemann}\footnote{This is the collection of his works where this paper can be found.} where he considered it as the expansion in hypergeometric series. 
The astronomical interest to investigate the gravitational potential of a torus appeared due to the discovery of the objects which consist of massive toroidal or ring structures whose gravitational fields can influence the dynamics around them: for instance, ring galaxies as the famous Hoag's object consisting of the central galaxy and the ring of star-formation around it \citep{1950AJ.....55Q.170H, 2011MNRAS.418.1834F}. Such a system has a peculiar dynamics due to the mutual gravitational forces from the central mass and the ring which act in opposite directions in the central region \citep{2018MNRAS.476.3269B}. Another example is the dusty geometrically-thick tori in active galactic nuclei (AGNs); they play an essential role in feeding the accretion disk and in providing the high luminosity of AGNs (see, e.g. \citep{2021agnf.book.....C}). N-body simulations of such toroidal structures consisting of clouds turn to be stable with an equilibrium cross-section of oval or elliptical shapes  \citep{2012MNRAS.424..820B, 2021MNRAS.503.1459B}.
Robust approximate expressions of the torus potential help to use them for the control and  interpretation of the N-body experiments. 
They can also be important for the magnetostatic and plasma physics including the tokamak design (see the references in \citep{2016AJ....152...35F}). 
Moreover the understanding of the properties of a torus potential can be helpful for the simulation of the gravitational field of asteroids with complicate shapes, as in the case of Benny asteroid, which has the ring-like solid mass distribution on its surface \citep{2020SciA....6.3350S}.
 The investigation of the torus potential also allow us to obtain the new results in potential theory in general.

The potential of the torus can be obtained by direct integration over the volume but the resulting expression is not convenient for the investigation of the physical properties of the torus. Another approach is to obtain the potential of this complex body considering it as the sum of more simple ones. This idea was used for the first time by I. Newton in the investigation of gravitational properties of a solid sphere using the spherical shells (see, for example, \citep{FCMbook}). Similar approach was used to obtain the potential of the homogeneous torus with a circular cross-section \citep{2011MNRAS.411..557B}. We used there a massive circle as the elementary component, which allowed us to discovery new gravitational properties of the torus. One of them is that the outer potential of the torus can be represented with good accuracy by the potential of a massive circle with the same mass located at the center of cross-section. 
 
The same method was used latter to obtain the potential of a thin toroidal shell  with the following investigation of this potential for the inner \citep{2019MNRAS.486.5656H} and outer  \citep{2020MNRAS.494.5825H} regions. Since the toroidal shell is a particular case of the solid torus, the representation of outer potential by that of the massive circle holds in this case too.  
\cite{2016AJ....152...35F} considered in detail the expansion in toroidal harmonics of the outer potential of the torus with different cross-section (oval, Brillouin toroid) and arbitrary density distribution.  The numerical experiments showed that the expansions converge rapidly, which saves  machine time. Indeed the investigation of the torus with cross-section different from circular is important for astrophysical problems because the presence the central mass and torus self-gravity leads to the elliptical/oval shapes of its cross-section.

In Section~\ref{Exact} we present the new expression for the gravitational potential of the torus with an elliptical cross-section. This was preliminary presented in \citep{2021MNRAS.503.1459B} in order to understand the result of N-body simulations.  In Section~\ref{TwoRings} we show that the outer potential of the elliptical torus can be represent by the potential of two massive circles. The striking result is that these massive circles are located halfway between the center and the foci of the elliptical cross-section.
	
\section{Integral expression for gravitational potential}
\label{Exact}
\begin{figure}
\centering
\includegraphics[width = 75mm]{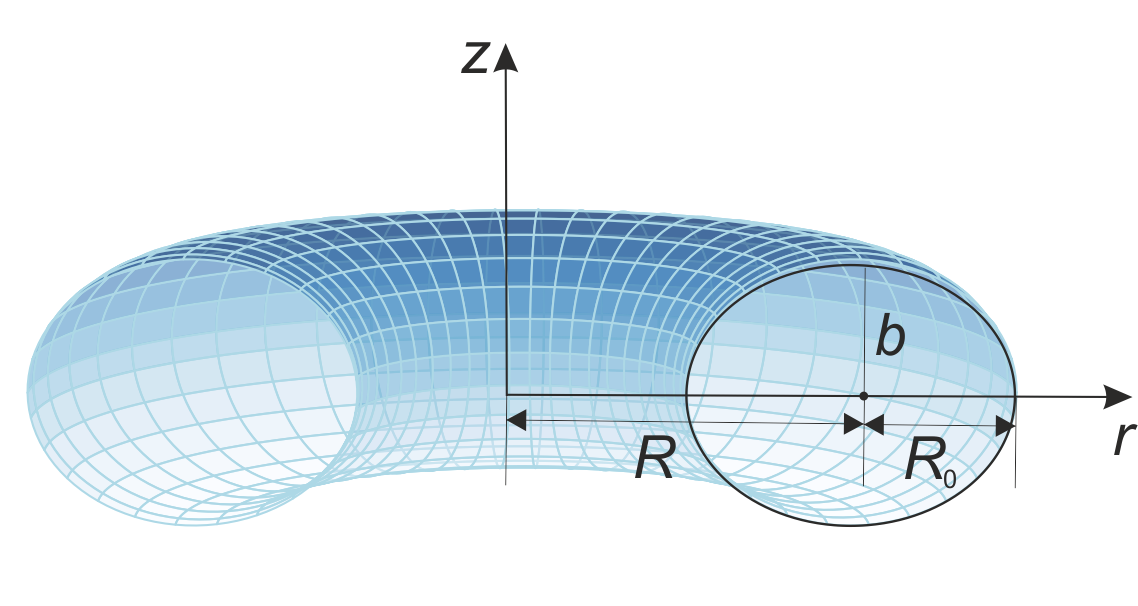}
\includegraphics[width = 70mm]{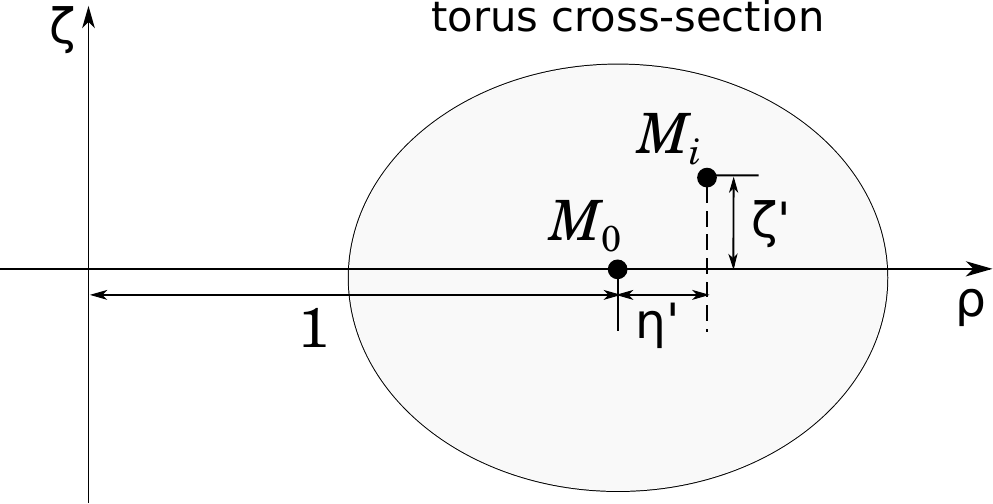}
\caption{Scheme of a torus with an elliptical cross-section.}
\label{fig:Ini}
\end{figure}

Consider a homogeneous torus with mass $M$,  major radius $R$, and an elliptical cross-section with the main semi-axis $R_0$ (Fig.~\ref{fig:Ini} {\it Top}). 
Since the torus is an axisymmetric body, we will consider the problem in the cylindrical coordinate system $(r,z)$ in which the equation of the elliptical cross-section is:
\begin{equation}\label{eq1}
\frac{(r-R)^2}{a^2}+\frac{z^2}{b^2}=1,
\end{equation}
where $a$, $b$, are major and minor semi-axis lengths of ellipse. Let us find the potential of the torus in an arbitrary point $P(r,z)$. It is convenient to use dimensionless coordinates
\mbox{$\rho = r/R$}, 
$\zeta = z/R$. For the following we will use the geometrical parameter $r_0 = R_0/R$,  which indicates the torus thickness. In this case  we will consider the dimensionless major, $a=r_0$, and minor, $b=\alpha r_0$, semi-axes of the cross-section, where the axes ratio is $\alpha=b/a$. 
So, the equation of the cross-section is:
\begin{equation}\label{eq2}
\frac{\eta^2}{r_0^2}+\frac{\zeta^2}{\alpha^2 r_0^2}=1,
\end{equation}
where the coordinate \mbox{$\eta=\rho -1$} is related to the reference system with the origin at the center of the cross-section.

To obtain the gravitational potential of the torus with the elliptical cross-section we made it consist of a set of massive circles\footnote{The term "massive circle" means the same as "infinitely thin ring" or "massive loop". };  we used the same approach for the circular cross-section case \citep{2011MNRAS.411..557B}.
The potential of the central massive circle with the mass $M_0$, which is located in the equatorial plane with the radius equal to the major radius of the torus $R$, has the form (see, for example, \citep{FCMbook}):
\begin{equation}\label{eq3a}
  \varphi_{mc,0}(\rho,\zeta) = \frac{GM_0}{\pi R} \, \phi_{mc,0}(\rho,\zeta),
\end{equation}
where the dimensionless potential is:
\begin{equation}\label{eq3b}
 \phi_{mc,0}(\rho,\zeta) = \sqrt{\frac{m}{\rho}} \, K(m),
\end{equation}
with the complete elliptical integral of the first kind: 
\begin{equation}\label{eq3c}
K(m)=\int_0^{\pi/2}\frac{d\beta}{\sqrt{1-m \sin^2 \beta}}
\end{equation}
and the parameter:
\begin{equation}\label{eq3d}
  m = \frac{4\rho}{(1+\rho)^2 + \zeta^2}~.
\end{equation}
To compose the torus of a set of the massive circles, we modify (\ref{eq3a}) for the case of the component circle with coordinates ($x', z'$) inside torus. They have  different radii $R'$ corresponding to the coordinates $x'$. We refer now the coordinate of the component massive circle to the center $C(R,0)$ of the torus cross-section: $x'= R' - R$.   
So, the potential of the component massive circle can be obtained by replacing  in (\ref{eq3a}) to (\ref{eq3d}): \mbox{$r/R =\rho$} with \mbox{$r/R'=r/(R+x')=\rho/(1+\eta')$} and $z/R=\zeta$ with \mbox{$(z-z')/R' = (\zeta - \zeta')/(1 + \eta')$}. Here we denote the coordinates of this circle (Fig.~\ref{fig:Ini},~{\it Bottom}): $\eta' = x'/R$ and $\zeta'= z'/R$.  As the result, the potential of component massive circle is:
\begin{multline}\label{eq3e0}
  \varphi_{mc,i}(\rho,\zeta; \eta',\zeta') = \frac{GM_i}{\pi R'}
  \sqrt{\frac{(1+\eta') \, m_i}{\rho}} \, K(m_i),
\end{multline}
where the parameter of the elliptical integral (\ref{eq3d}) $m$ transforms to $m_i$ as:
\begin{multline}\label{eq3g}
  \displaystyle m_i = \frac{4 r/R'}{(1+r/R')^2 +(\zeta-\zeta')^2/R'^2} = \\
   = \frac{4\rho \, (1+\eta')}{(1+\eta'+\rho)^2 +
  (\zeta-\zeta')^2}~.
\end{multline}
The homogeneity of the torus implies the equality of the reduced masses $\kappa_0 = \kappa_i$,
where $\kappa_0 = M_0/(2\pi R)$ is the reduced mass of the central circle and $\kappa_i = M_i/(2\pi R')$. Then we have  $M_i = M_0 R'/R$, and substituting it in (\ref{eq3e0}) we have:
\begin{equation}\label{eq3e}
  \varphi_{mc,i}(\rho,\zeta; \eta',\zeta') = \frac{GM_0}{\pi R} \, \phi_{mc,i}(\rho,\zeta; \eta',\zeta'),
\end{equation}
where its dimensionless potential:
\begin{equation}\label{eq3f}
  \phi_{mc,i}(\rho,\zeta; \eta',\zeta') = \sqrt{\frac{(1+\eta') \,
  m_i}{\rho}} \, K(m_i).
\end{equation}
For the central circle $\eta'=0$, $\zeta'=0$ and (\ref{eq3g}) to (\ref{eq3f}) pass to the expressions (\ref{eq3a}) to (\ref{eq3d}).

We find the torus potential integrating over the elliptical cross-section with the boundaries determined by (\ref{eq2}).
For this we change the mass $M_0$ in (\ref{eq3e}) with the differential one:
\begin{equation}\label{eq9}
  dM=\frac{M}{\pi \alpha \, r_0^2} d\eta'd\zeta',
\end{equation}
where $M$ is the total mass of all massive circles and it is equal to the torus mass.
The final expression of the potential of homogeneous torus with the elliptical cross-section is:
\begin{multline}\label{eq10}
\varphi^\text{ell}_\text{torus} (\rho, \zeta) = \frac{G M}{\alpha \pi^2 r_0^2 R} 
	\times \\ 
	\int_{-r_0}^{r_0} 
        \int_{-\alpha \sqrt{r_0^2 - \eta'^2}}^{\alpha \sqrt{r_0^2 - \eta'^2}}
            d\eta'd\zeta' \phi_{mc,i}(\rho,\zeta; \eta', \zeta'),
\end{multline} 
where $\phi_{mc,i}$ is determined by (\ref{eq3f}) and the parameter $m_i$ by (\ref{eq3g}).
This expression (\ref{eq10}) holds for any point of space, both inside and outside of the torus volume. For the limited case $\alpha=1$ we obtain the potential of the torus with the circular cross-section. 
\begin{figure}
\centering
\includegraphics[width = 70mm]{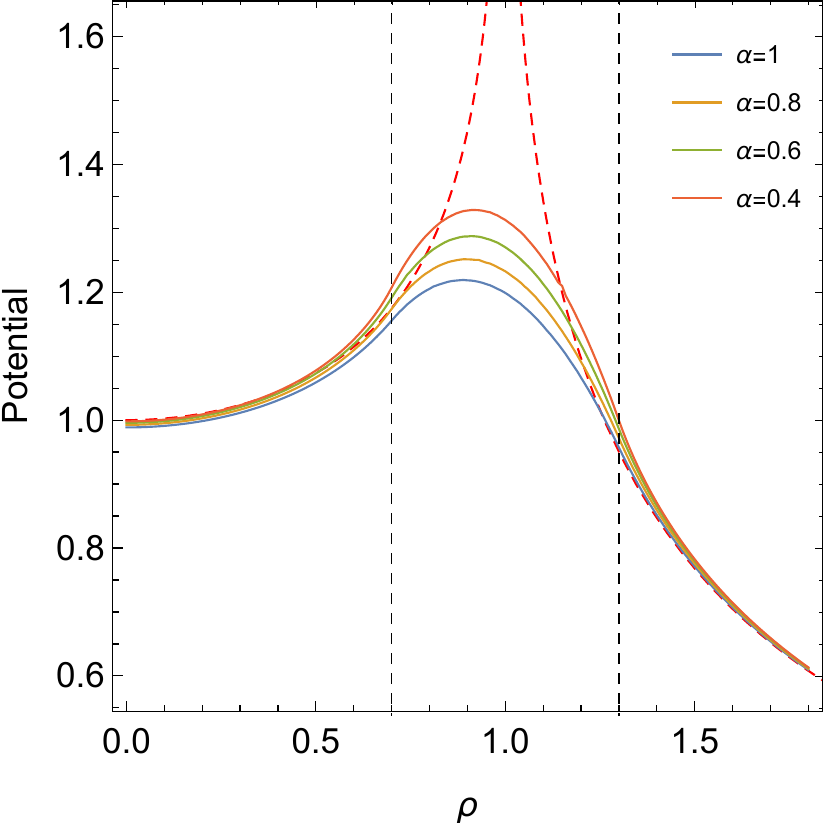} \qquad
\includegraphics[width = 70mm]{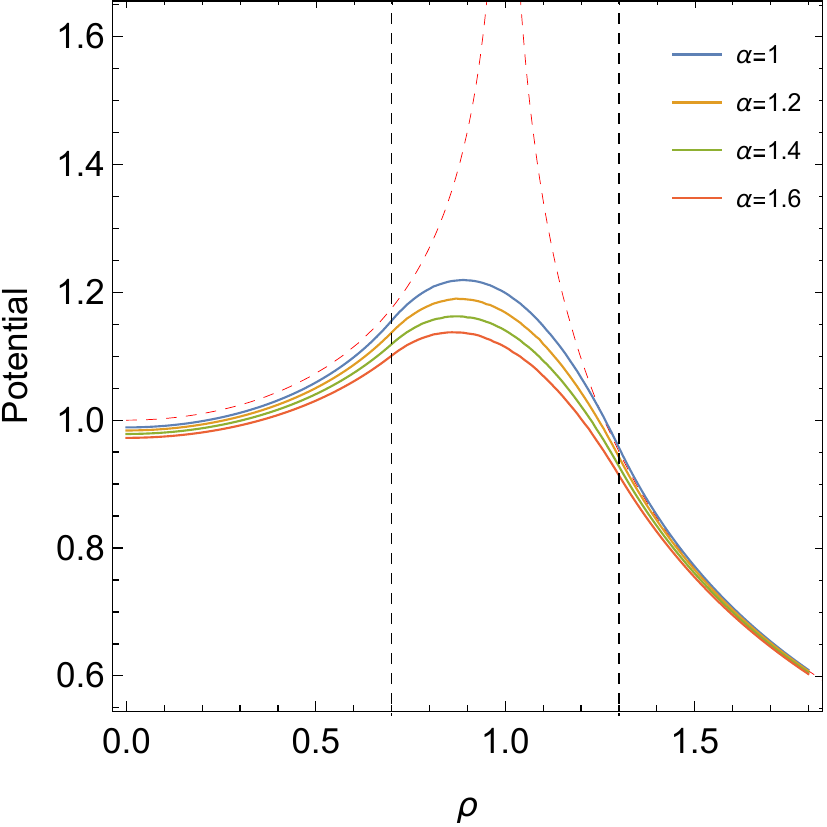}
\caption{Potential curves $\varphi^\text{ell}_\text{torus} (\rho, 0)$
for tori with  the same geometrical parameter \mbox{$r_0=0.3$} and different values 
of the axis ratio: $\alpha\leq 1$ ({\it top}), $\alpha\geq 1$ ({\it bottom}).
The potential of the massive circle $\varphi_{mc,0}$ with the same mass $M$ and radius $R$ is marked by red dashed line. Black dashed lines mark the boundary of the torus surface. }
\label{fig:CurvesAlpha}
\end{figure}

Fig.~\ref{fig:CurvesAlpha} shows the curves of the gravitational potential of the torus for a chosen value of the geometrical parameter $r_0$ and for different axis ratios $\alpha$ of the cross-section\footnote{For all numerical simulations we use the unit system \mbox{$M=R=G=1$}.}. These curves correspond to the case when the point $P$ is in the equatorial plane ($\zeta=0$). 
It can be seen that the maximum of the potential curves of the flattened torus (oblate cross-section) increase. This is related to the fact that we fix the major semi-axis of the cross-section and that the more flattened torus contains more mass in the main equatorial plane (Fig.~\ref{fig:CurvesAlpha}, {\it Top}). For the torus with an elongated (prolate) cross-section (Fig.~\ref{fig:CurvesAlpha}, {\it Bottom}), all curves are lower than that the curve of the potential for the central massive circle with mass equal to the torus mass $M$ and with the radius equal to the major torus radius $R$. This results show that the outer potential of the elliptical torus can be represented with the massive circle potential with some modification. We can predict, on the basis of our previous results for the circular torus, that this modification should be some function of the parameters $r_0$ and $\alpha$ which we will proof in the paper II. In the next Section we will show that two massive circles can represent the outer potential of the elliptical torus with good accuracy if these circles are located at some distances from the center of the cross-section.     

\section{Representation of the outer potential of the torus by two massive circles }
\label{TwoRings}

In the following  we call outer region of the torus the region outside the torus body where the potential satisfies Laplace equation. It means that the coordinates of the point $P(\rho, \zeta)$ satisfy the condition:
\begin{equation}\label{eq4.00}
\alpha^2(\rho - 1)^2 + \zeta^2 \geq \alpha^2 r_0^2.
\end{equation}
As it was remarked above, the outer potential of a homogeneous torus with a circular cross-section can be represented with a good accuracy by a massive circle placed at the cross-section center. By analogy we can guess that the outer potential of the torus with an elliptical cross-section can be represented by two massive circles located at some equal distance\footnote{We made a set of numerical experiments for the case when the distances to the massive circles are not equal to each other. As a result we understood that the best solution corresponds to the case of equal distances and equal masses for the two circles.} from torus cross-section center in opposite directions along the major axis. 
To test this idea we use the expression for the component circle (\ref{eq3e}) with $i=1,2$; the coordinates of these two circles are: $(\eta'_{1,2}, \zeta'_{1,2}) = (l_{1,2},0) = (\pm l,0)$.
In this case we can write the outer potential of the torus as: 
\begin{equation}\label{eq4.0}
 \varphi^\text{ell}_\text{torus}(\alpha, r_0) \approx
 \Big(\varphi_{mc,1}(l,0) + \varphi_{mc,2}(-l,0)\Big) g(r_0, \alpha),
\end{equation} 
where $l=l(\alpha, r_0)$ is the distance from the cross-section center which, in the general case, depends on the axis ratio $\alpha$ and the geometrical parameter $r_0$.
In the brackets of potential functions we have highlighted just the parameters, but obviously these functions depend also on the coordinates $(\rho, \zeta)$. 
The function $g(r_0, \alpha)$ can be determined by the limiting case of the torus with the circular cross-section ($\alpha =1$) that is the S-approximation obtained in \citep{2011MNRAS.411..557B}:
\begin{equation}\label{eq3.8}
\varphi^\text{circ}_\text{torus}(\rho, \zeta; r_0) \approx \frac{GM}{\pi R}\phi_{mc,0}\left( 1-\frac{r^2_0}{16} + \frac{r^2_0}{16}  \, S(\rho, \zeta)\right)~,
\end{equation}
where:
\begin{equation}\label{eq3.9}
S(\rho, \zeta) = \frac{\rho^2 + \zeta^2 - 1}{(\rho - 1)^2 + \zeta^2}\frac{E(m)}{K(m)},
\end{equation}
and $E(m)$ is the complete elliptical integral of the second kind:
\begin{equation}\label{eq3.9a}
E(m) = \int_0^{\pi/2}d\beta\sqrt{1-m \sin^2 \beta},
\end{equation}
with the parameter $m$ determined by (\ref{eq3d}).
In this case the two circles merge and $\varphi^\text{ell}_\text{torus} \rightarrow  \varphi^\text{circ}_\text{torus}$.
In order to satisfy the limiting case of a circular torus and account for an elliptical cross-section, we may generalize the function $g$ as:
\begin{equation}\label{eq3.9l}
g(r_0,\alpha) = \left( 1-\alpha^2\frac{r^2_0}{16} + \alpha^2\frac{r^2_0}{16} \, S(\rho, \zeta)\right).
\end{equation} 
In the expression for the component circle (\ref{eq3e}) the mass of each of the two circles is set equal to half the torus mass, $M_0=M/2$. 

We now want to estimate the "best" value for the distance $l$.
To this end we minimize the average differences between the results with the approximate expression for the outer potential (\ref{eq4.0}) and those with the exact one (\ref{eq10}) in a region which is indicated in Figs \ref{fig:Isoph09}--\ref{fig:Isoph13} (note that at larger distances from the torus surface the agreement improves).
We express the distance $l$ as function of the focal distance $f$ and some constant parameter $k$ in the form: $l=f/(k + \alpha)$. Then we made the set of simulations fixing $r_0$, $\alpha$ and varying $k$, and compared the approximate potential values in each point with the exact integral expression. As the result we obtain the maps of the relative errors, RE:
\begin{equation}\label{eq3.9b}
\text{RE}(\rho_i, \zeta_j; r_0,\alpha, k) = 1- \frac{\varphi_{app}(\rho_i, \zeta_j; r_0,\alpha, k)}{\varphi_{ex}(\rho_i, \zeta_j; r_0,\alpha)},
\end{equation}
where $\varphi_{ex}$ is the integrate expression 
(\ref{eq10}) and $\varphi_{app}$ is the potential represented by two massive circles (\ref{eq4.0}); \mbox{$i,j=1 \dots N$}. The number of coordinates for all simulations: $N \times N = 120 \times 120$.
The criterion to chose the "best" value for $k$ is that the mean absolute error and standard deviation are minimal. For all cases we discovered that $k+\alpha = 2$ gives the best coincidence for all the values: $0.4<\alpha \le 1$ and $0<r_0\le 0.8$ for the oblate case and $1\le\alpha < 1.6$ and $0<r_0\le 0.7$ for the prolate case. It means that  value of $l$ is at half of the distance to the focus (Fig.~\ref{fig:ellipse}). We demonstrate it in  Appendix~\ref{App}.

Finally, the approximate expression of the outer potential of the homogeneous torus with an elliptical cross-section is:
\begin{equation}\label{eq4.1}
 \varphi^\text{ell}_\text{torus}(\rho,\zeta) \approx \frac{GM}{2\pi R}
 (\phi_{mc,1} + \phi_{mc,2})\Big(1-\alpha^2 \frac{r^2_0}{16} + \alpha^2 \frac{r^2_0}{16} \, S \Big),
\end{equation}  
where $\phi_{mc,1,2} = (\phi_{mc,1}, \phi_{mc,2})$ are the dimensionless potentials of two massive circles:
\begin{multline}\label{eq4.3}
\phi_{1,2} = \sqrt{\frac{m_{1,2}(1 \pm l)}{\rho}} \, K(m_{1,2}) = \\ 
= \frac{2(1 \pm l)}{\sqrt{(\rho \pm l +1)^2 + \zeta^2}} \, K(m_{1,2}),
\end{multline}
with the corresponding parameters $m_{1,2}=(m_1, m_2)$:
\begin{equation}\label{eq4.4}
m_{1,2} = \frac{4\rho \, (1 \pm l)}{(\rho \pm l + 1)^2 + \zeta^2},
\end{equation} 
and function $S$ is determined by (\ref{eq3.9}).
The massive circles are located at the equal distance $(l_1, l_2) =(+l, -l)$ from the center of the torus cross-section:
\begin{equation}\label{eq4.5}
 l(\alpha, r_0) =  \frac{f(\alpha, r_0)}{2} = \frac{r_0\sqrt{1-\alpha^2}}{2}, 
\end{equation} 
where $f$ is the distance to the focus.

\begin{figure}
\includegraphics[width =40mm]{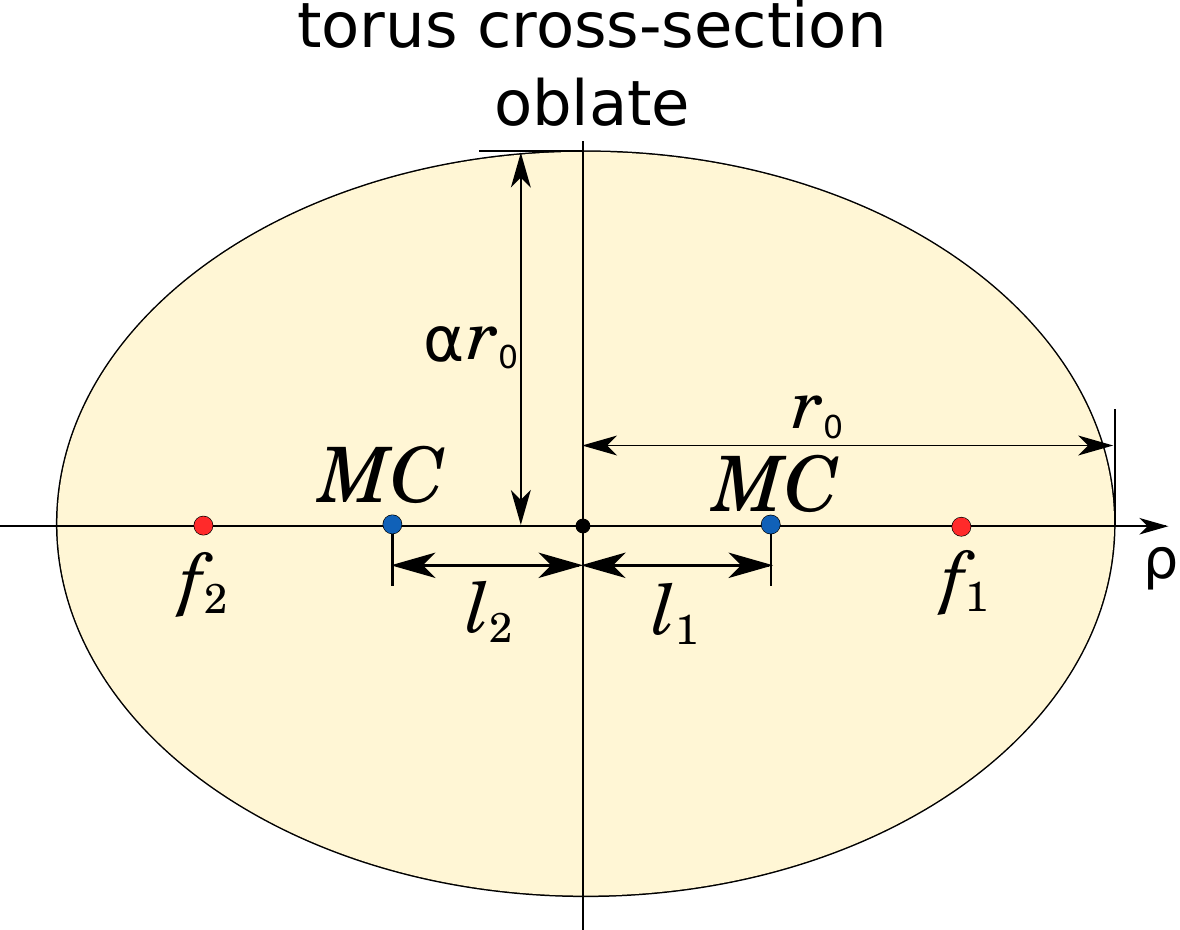}
\includegraphics[width =40mm]{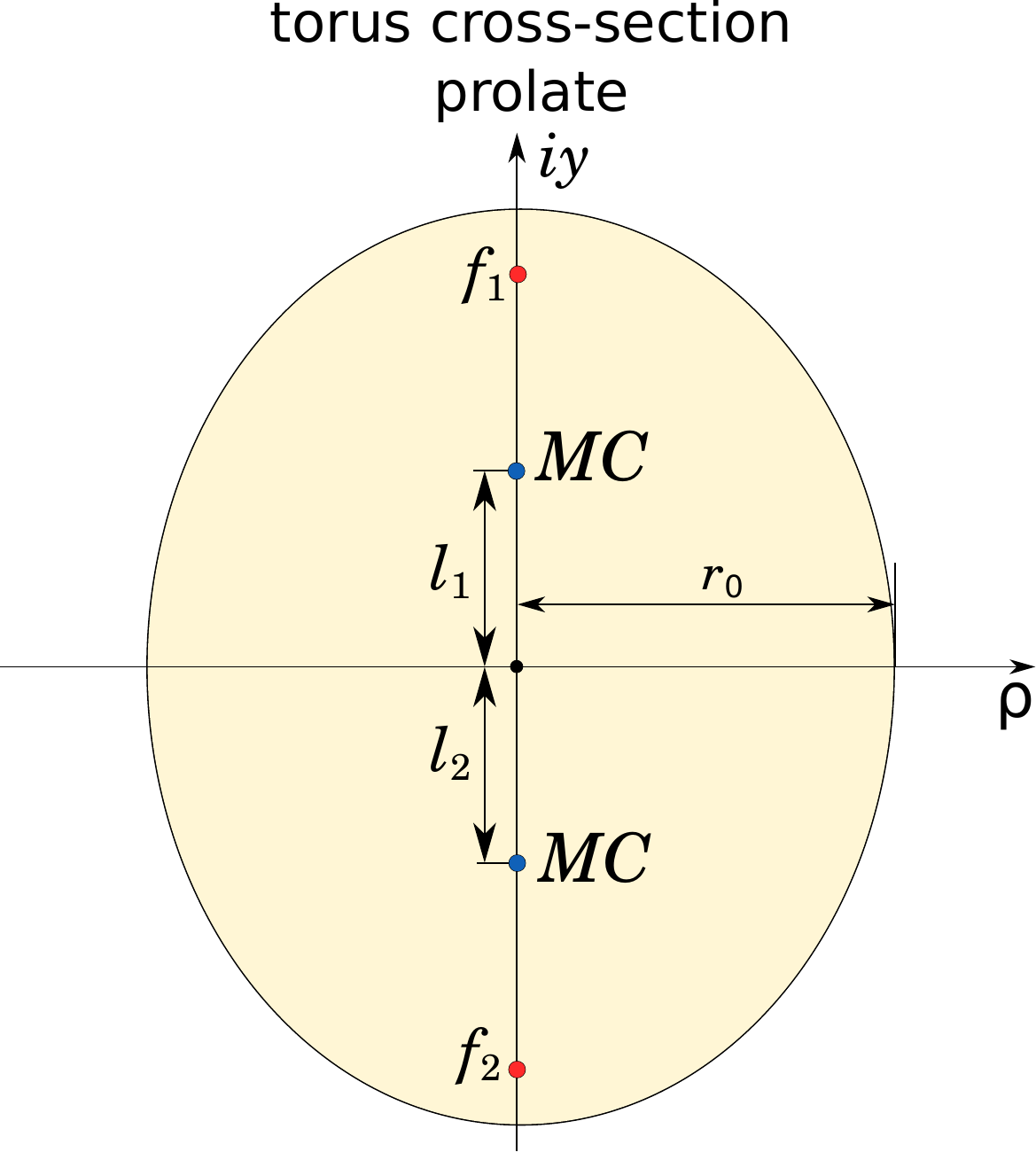}
\caption{Sketch of the torus cross-section and locations of the massive circles (MCs) for the oblate ({\it left}) and  prolate ({\it right}) cases.}
 \label{fig:ellipse}
\end{figure}

Since we are using as the limiting case the S-approximation for the torus with circular cross-section (\ref{eq3.8}),
we can use the simplifications presented in \citep{2011MNRAS.411..557B}. So, we have: 
\begin{equation}\label{eq4.1bb}
 \varphi^\text{ell}_\text{torus}(\rho,\zeta) \approx \frac{GM}{2\pi R}
 (\phi_{mc,1} + \phi_{mc,2})\Big(1-\alpha^2 \frac{r^2_0}{16} + \alpha^2 \frac{r^2_0}{16} \frac{\zeta^2 - 1}{\zeta^2 + 1} \Big).
\end{equation}  
This expression is obtained for the case when $\rho \rightarrow 0$ because it leads to $m\rightarrow 0$ and the ratio of the elliptical integrals $E(m)/K(m)\rightarrow 1$ in (\ref{eq3.9}). For the equatorial plane the expression (\ref{eq4.1bb}) becomes more simple:
\begin{equation}\label{eq4.1cc}
 \varphi^\text{ell}_\text{torus}(\rho,0) \approx \frac{GM}{2\pi R}
 (\phi_{mc,1} + \phi_{mc,2})\Big(1-\alpha^2 \frac{r^2_0}{8} \Big).
\end{equation} 
In the symmetry point of the torus, \mbox{$\rho=\zeta=0$}, the dimensionless potential of the massive circles is \mbox{$\phi_{mc,1} = \phi_{mc,2} = 2K(0) = \pi$} and we have:
\begin{equation}\label{eq4.dd}
 \varphi^\text{ell}_\text{torus}(0,0) \approx \frac{GM}{   R}
 \Big(1-\alpha^2 \frac{r^2_0}{8} \Big).
\end{equation}
We can see that in this point the potential depends on the axis ratio and geometrical parameter. The condition $\varphi^\text{ell}_\text{torus}(0,0)>0$ is satisfied always for all $\alpha$ since $\alpha < 2\sqrt{2}/r_0$ for the possible values  $0 < r_0 < 1$.     
The expressions (\ref{eq4.1bb}) and (\ref{eq4.1cc}) can be useful for analytical investigation because they are expressed in a more simple form: for example, in the dynamical problems requiring the use of the force components in the equations of motion. 

\subsection{Case of torus oblate cross-section}
\begin{figure*}
\includegraphics[width =50mm]{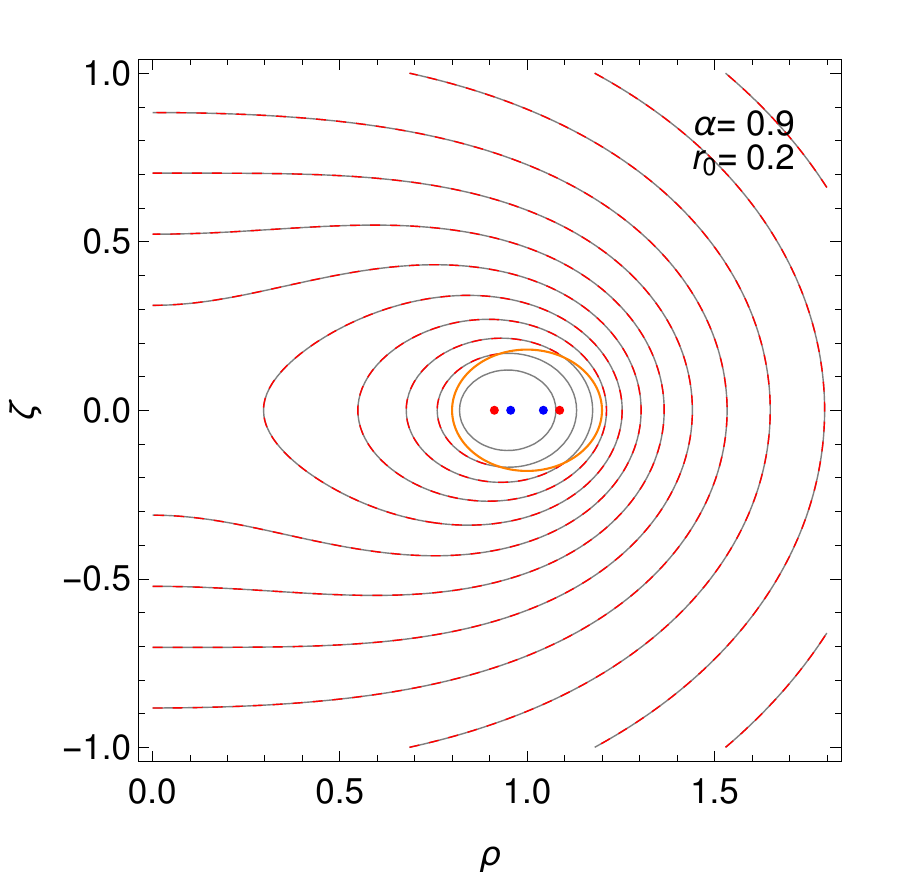}\qquad\qquad
\includegraphics[width =65mm]{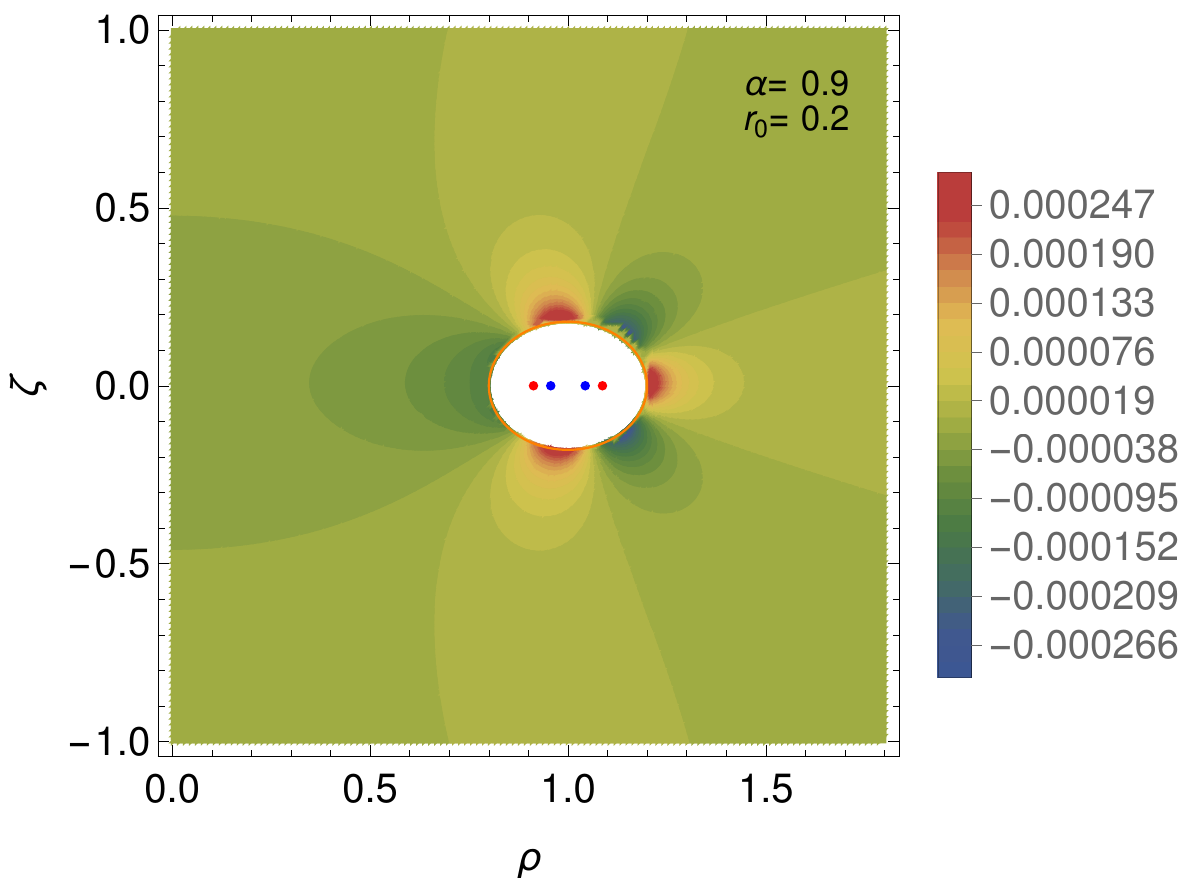}
\includegraphics[width =50mm]{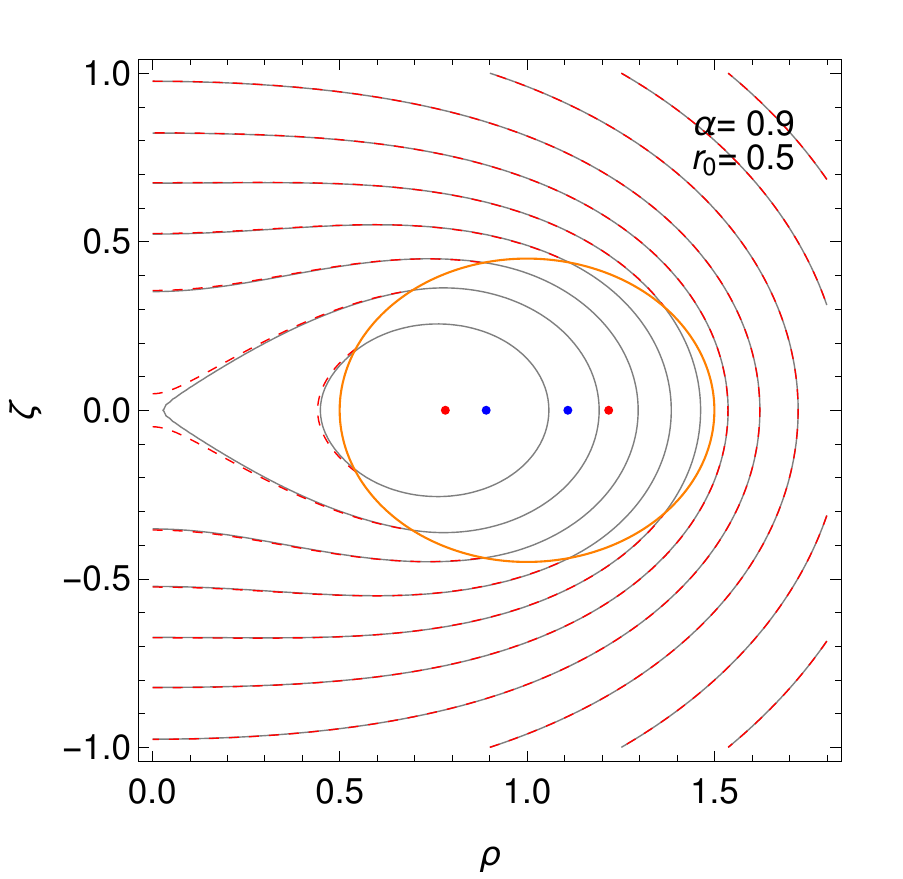}\qquad\qquad 
\includegraphics[width =65mm]{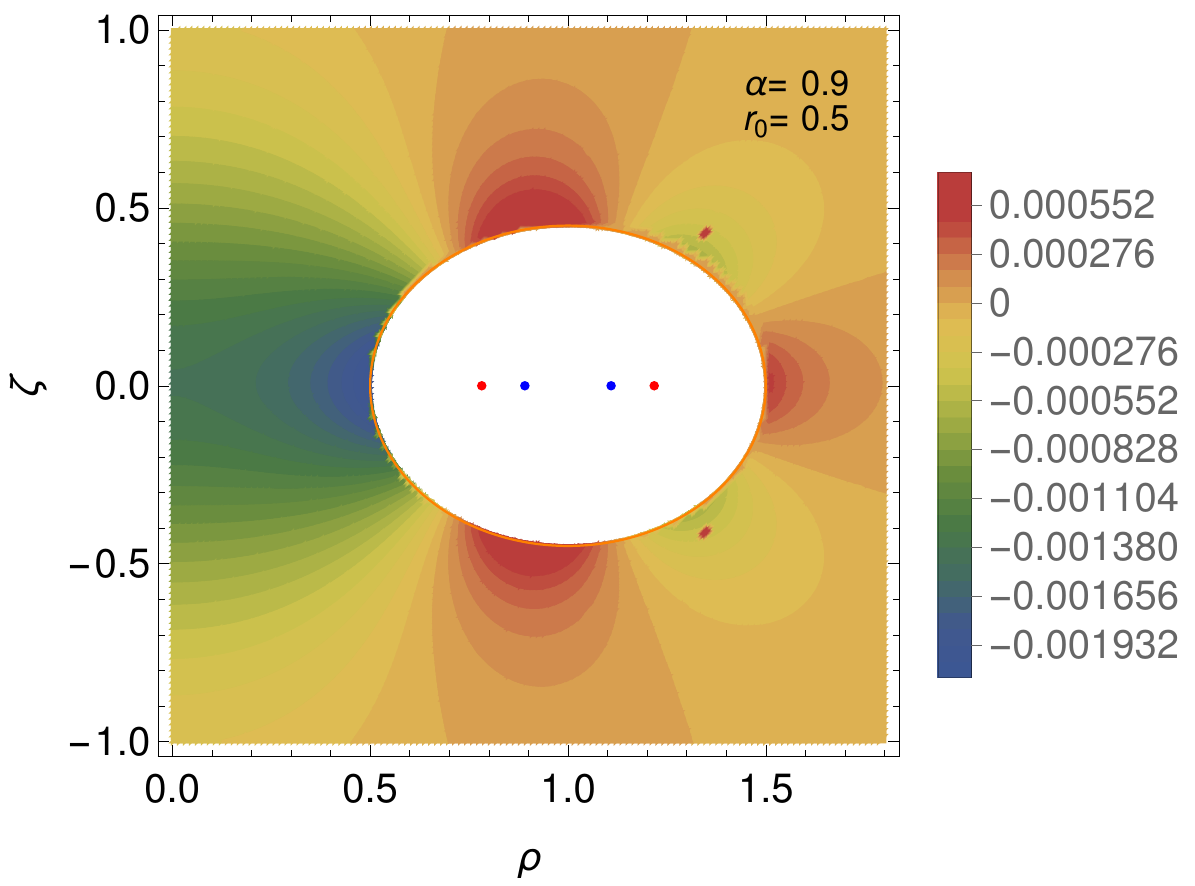}
\caption{{\it Left}: Isopotential curves corresponding to the exact integral expression (\ref{eq10})  ({\it black curves}) and to the approximation by two massive circles (\ref{eq4.1}) ({\it dashed red curves}) of the oblate torus with $\alpha = 0.9$, $r_0 = 0.2, 0.5$. The cross-section boundary is marked by the orange curve with its foci ({\it red points}). The positions where the massive circles intercept the cross-section are marketed by blue points. {\it Right}: Relative error maps.}
 \label{fig:Isoph09}
\end{figure*}
\begin{figure*}
\includegraphics[width =50mm]{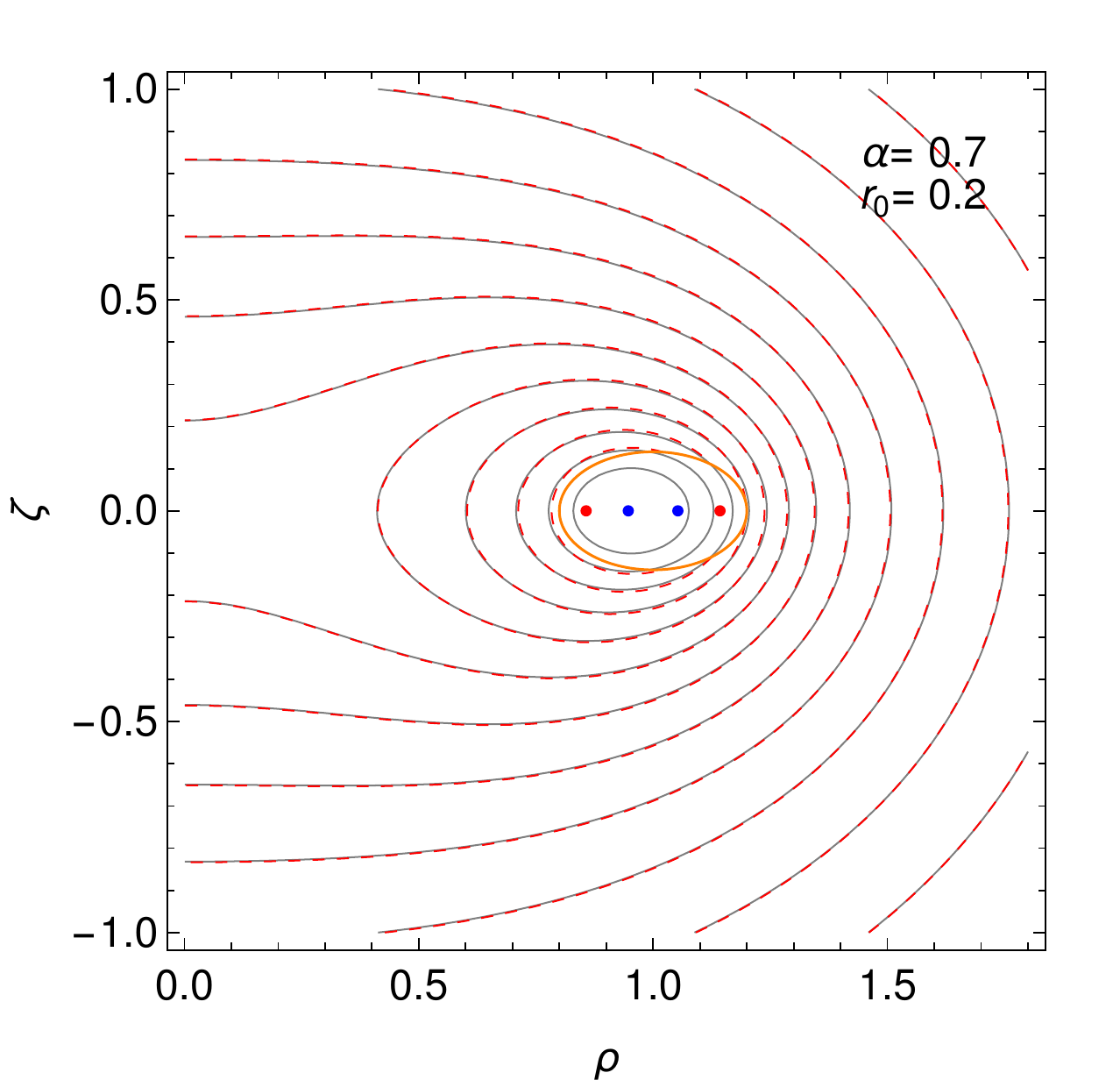}\qquad\qquad
\includegraphics[width =65mm]{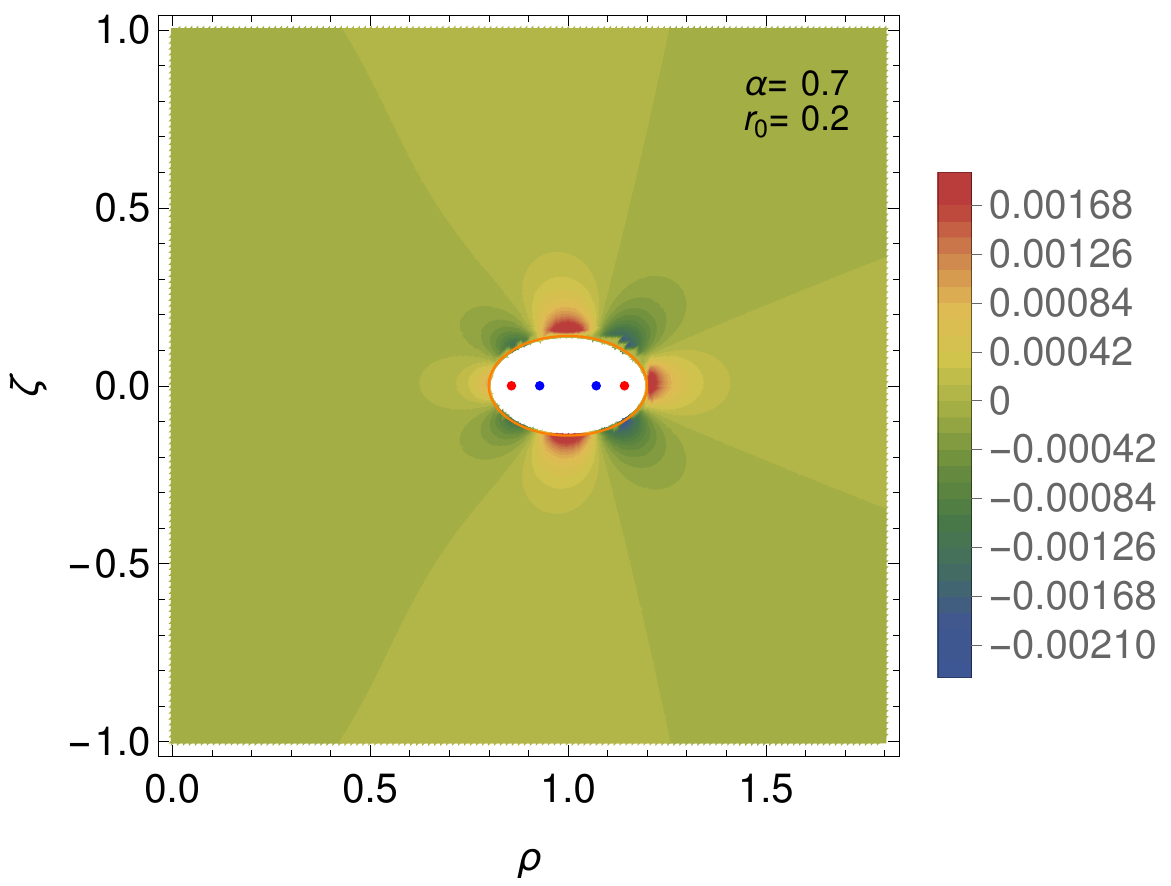}
\includegraphics[width =50mm]{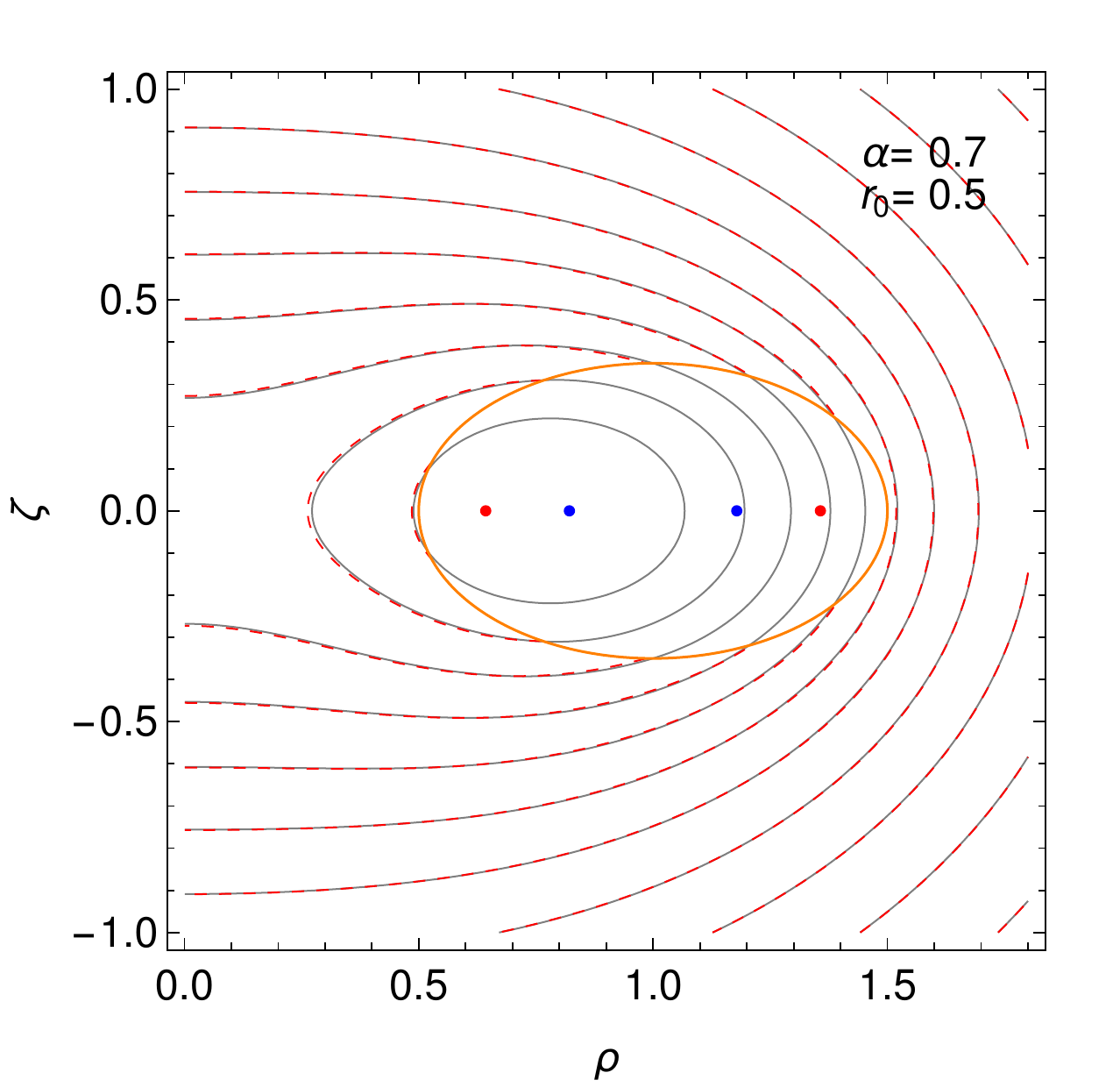}\qquad\qquad 
\includegraphics[width =65mm]{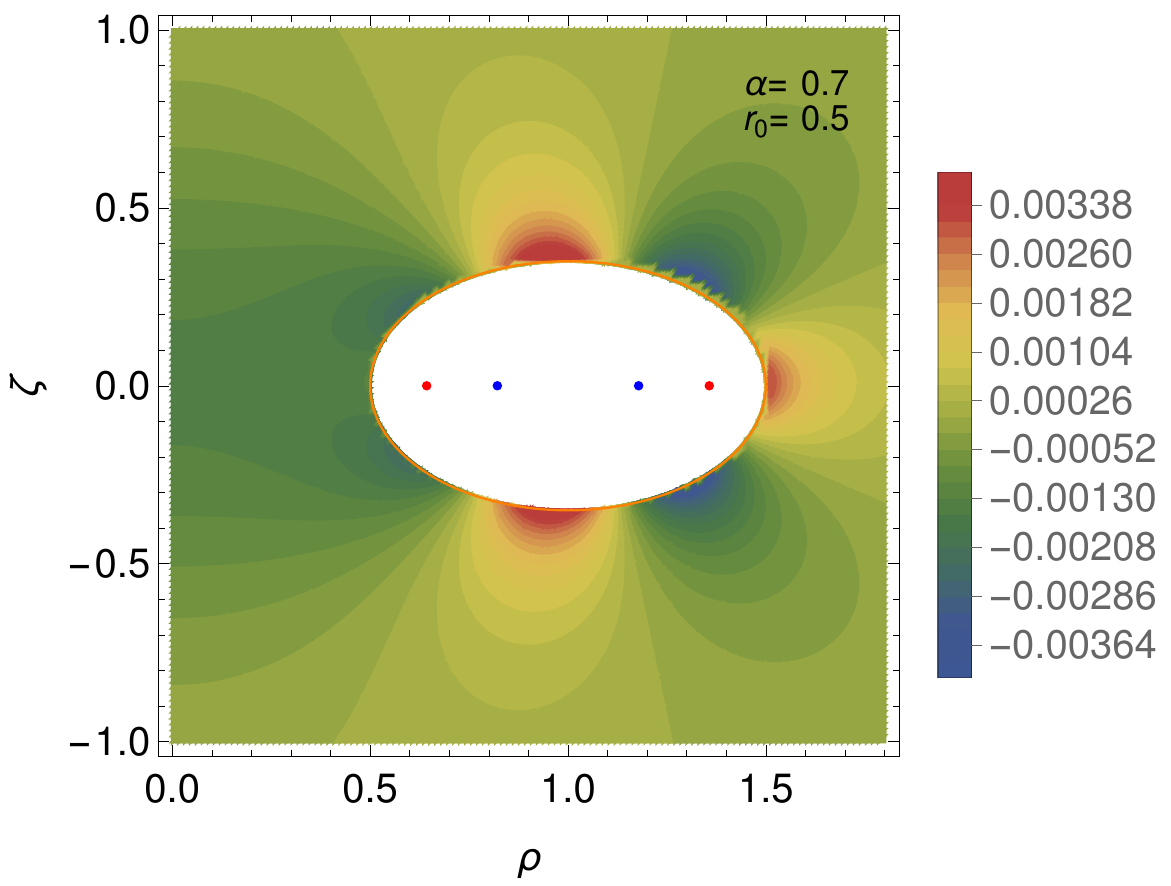}
 \caption{Same as Fig. \ref{fig:Isoph09} but for $\alpha = 0.7$.}
 \label{fig:Isoph07}
\end{figure*}

Examples of relative error maps calculated by the formula (\ref{eq3.9b}) but with the final approximate expression of the outer potential (\ref{eq4.1}) are shown in Figs.~\ref{fig:Isoph09} and \ref{fig:Isoph07} for the oblate case. It is seen that the representation of outer torus potential by two massive circles is really robust even for fairly flattened torus. For the low  value of the torus cross-section axis ratio $\alpha= 0.4$ and for the large value of $r_0 = 0.8$, the mean absolute error is about 0.5\% and the mean standard deviation about 0.7\%. 
This is an extreme cases; in all the others where the torus is either thinner and/or with rounder cross section, the agreement improves.
For the smaller values of $\alpha$, instead, the potential profiles have higher errors near the torus surface; but in this case the torus degenerates into a disk with a hole and the potential of a disk can be used.        

\subsection{Case of torus prolate cross-section}
\begin{figure*} 
\includegraphics[width =50mm]{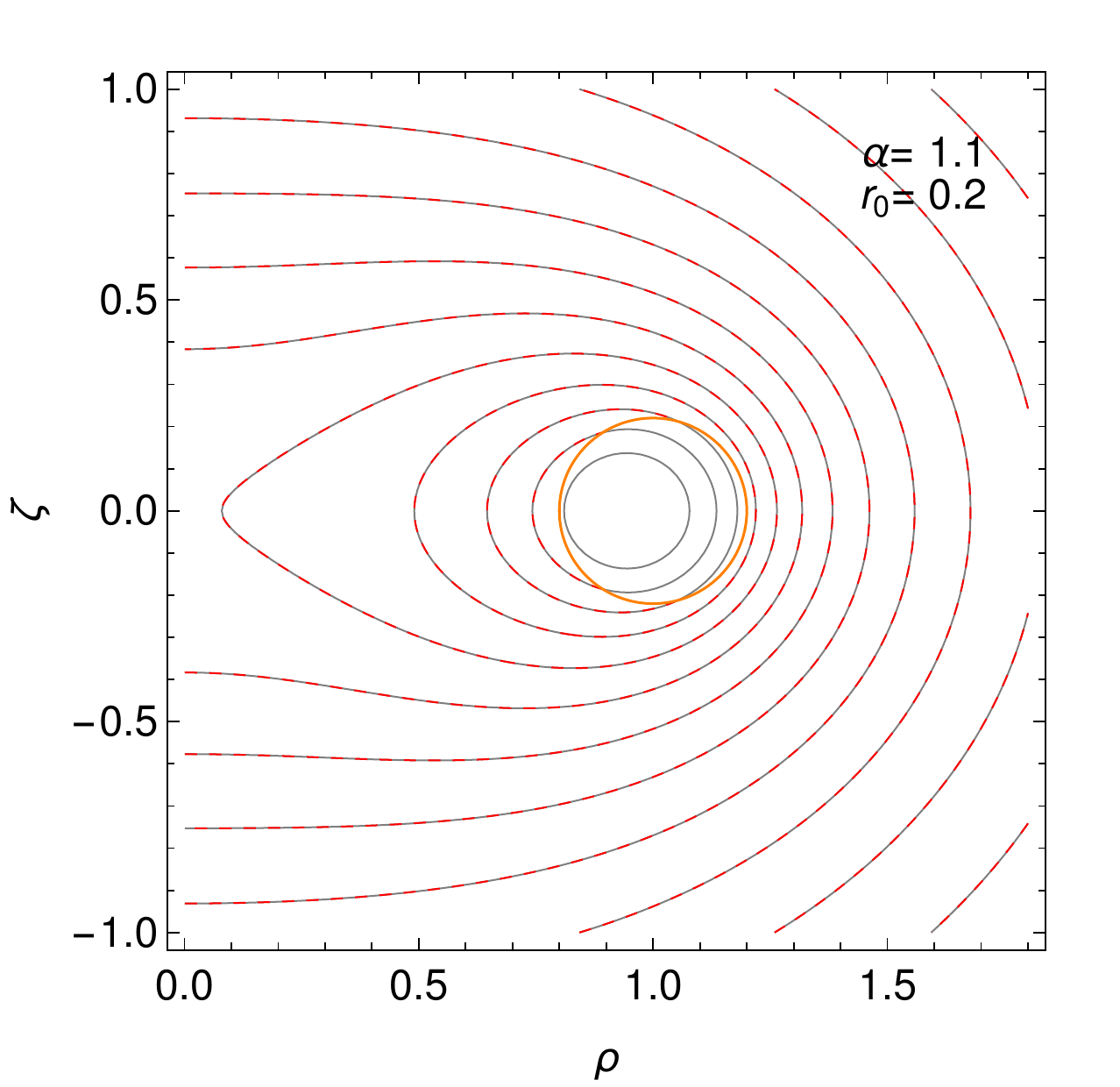}\qquad\qquad 
\includegraphics[width =65mm]{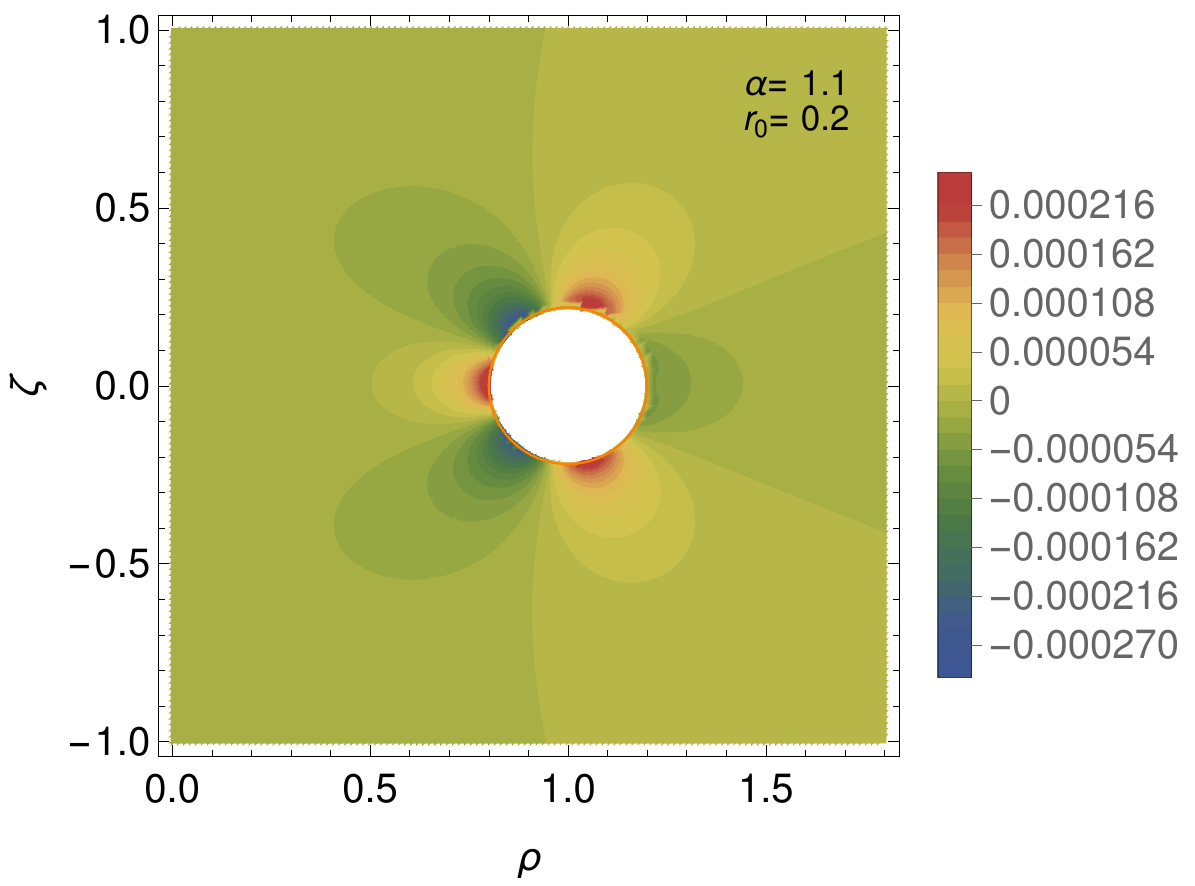}
\includegraphics[width =50mm]{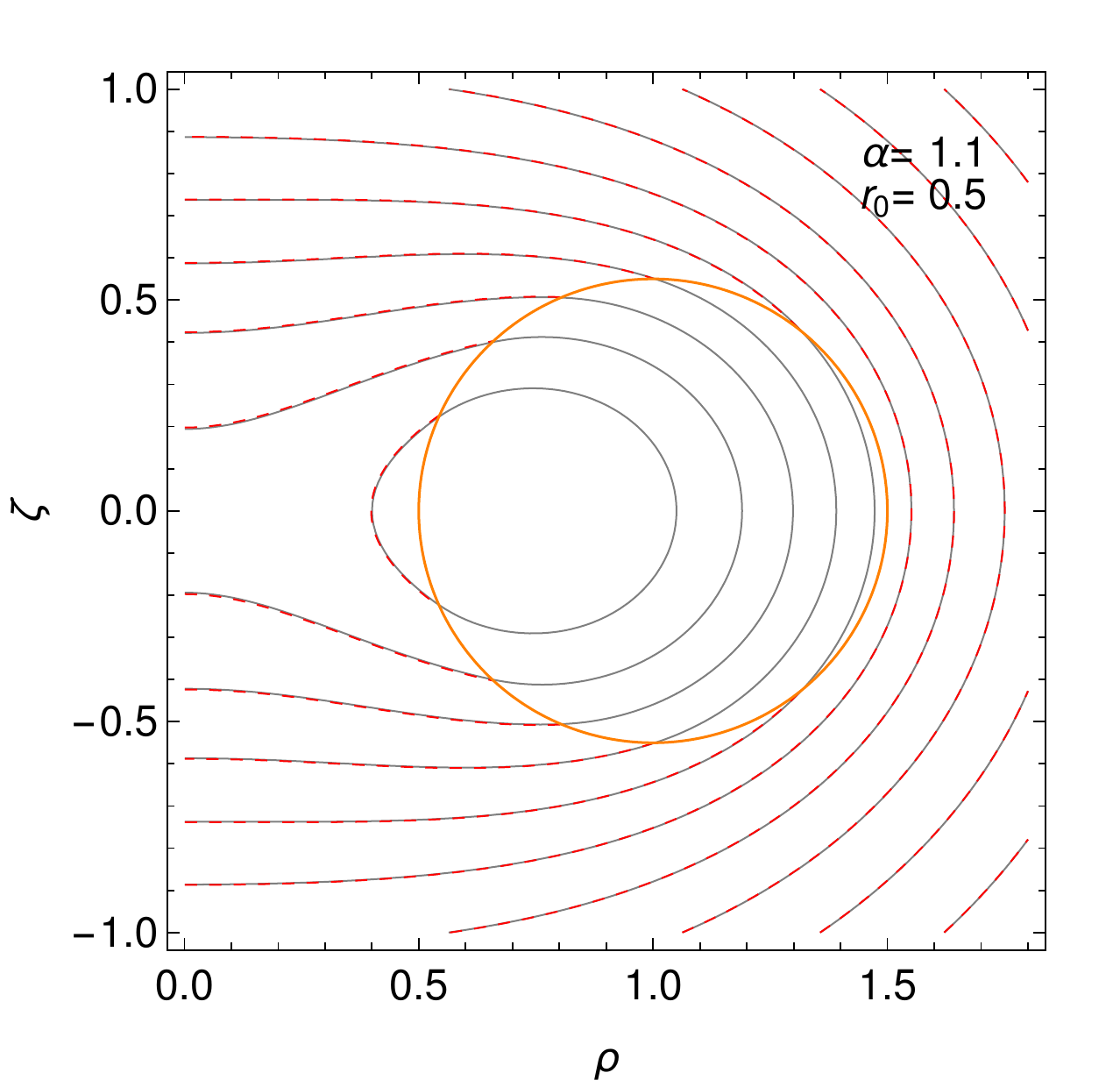}\qquad\qquad 
\includegraphics[width =65mm]{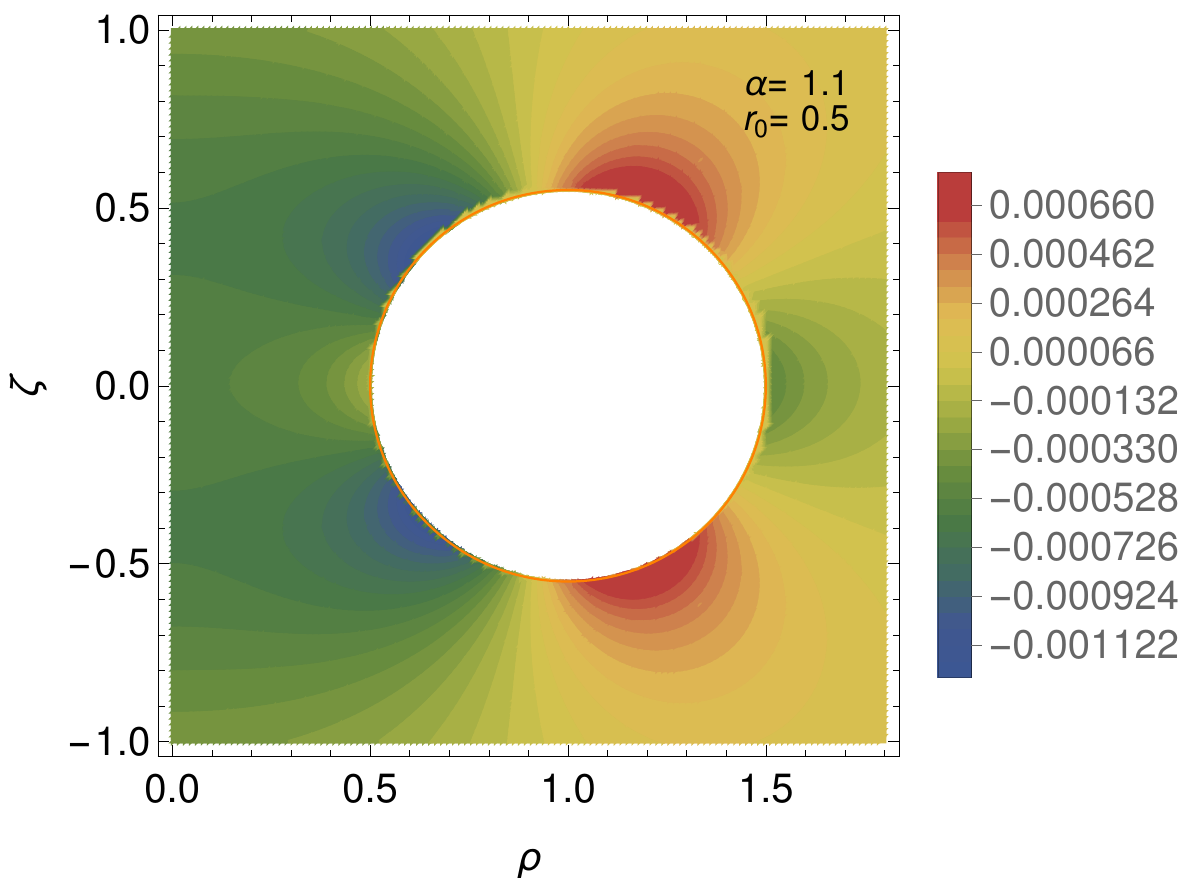}
 \caption{The same that on Fig.\ref{fig:Isoph09} but for $\alpha = 1.1$.}
 \label{fig:Isoph11}
\end{figure*}

\begin{figure*} 
\includegraphics[width =50mm]{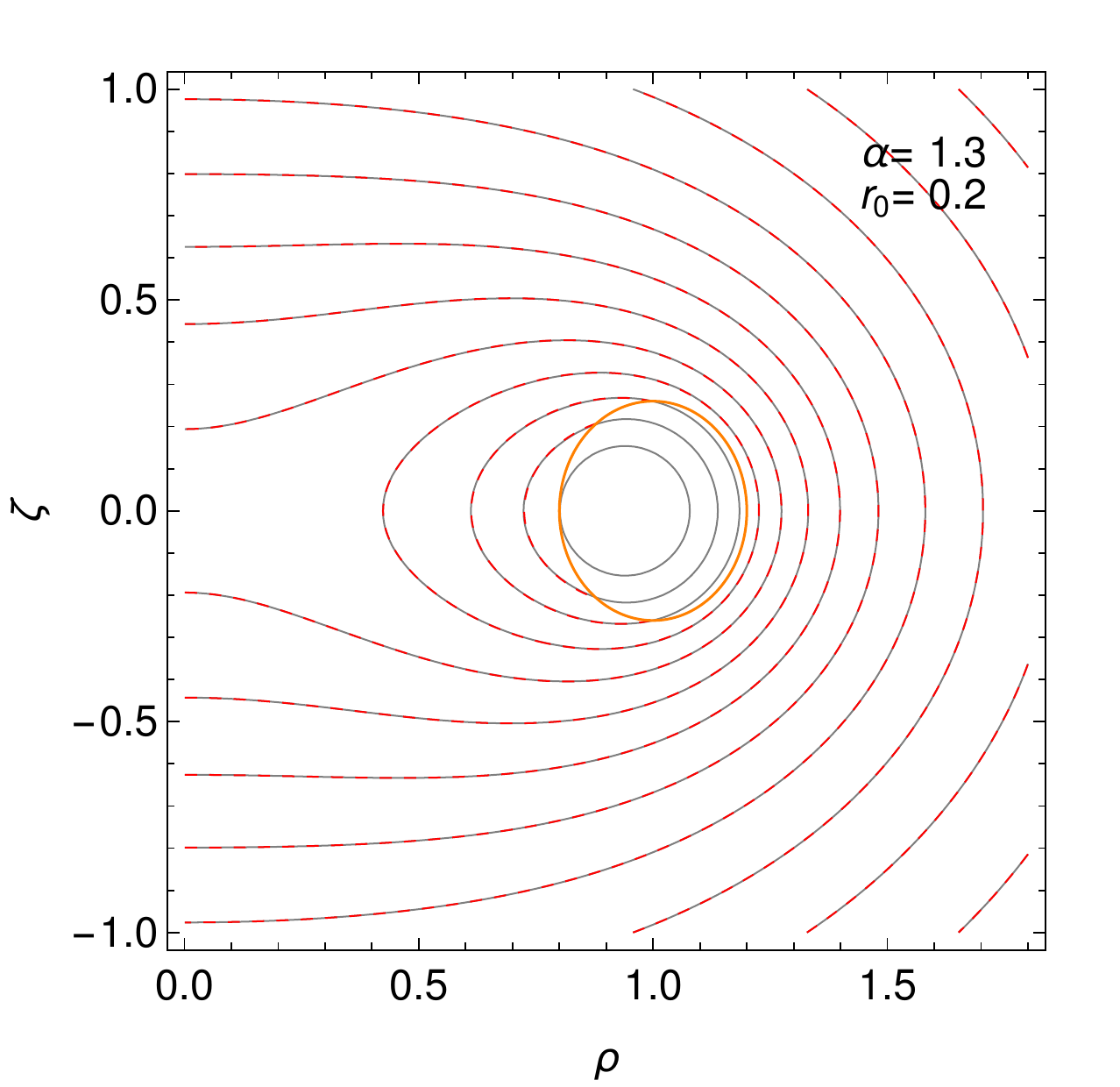}\qquad\qquad 
\includegraphics[width =65mm]{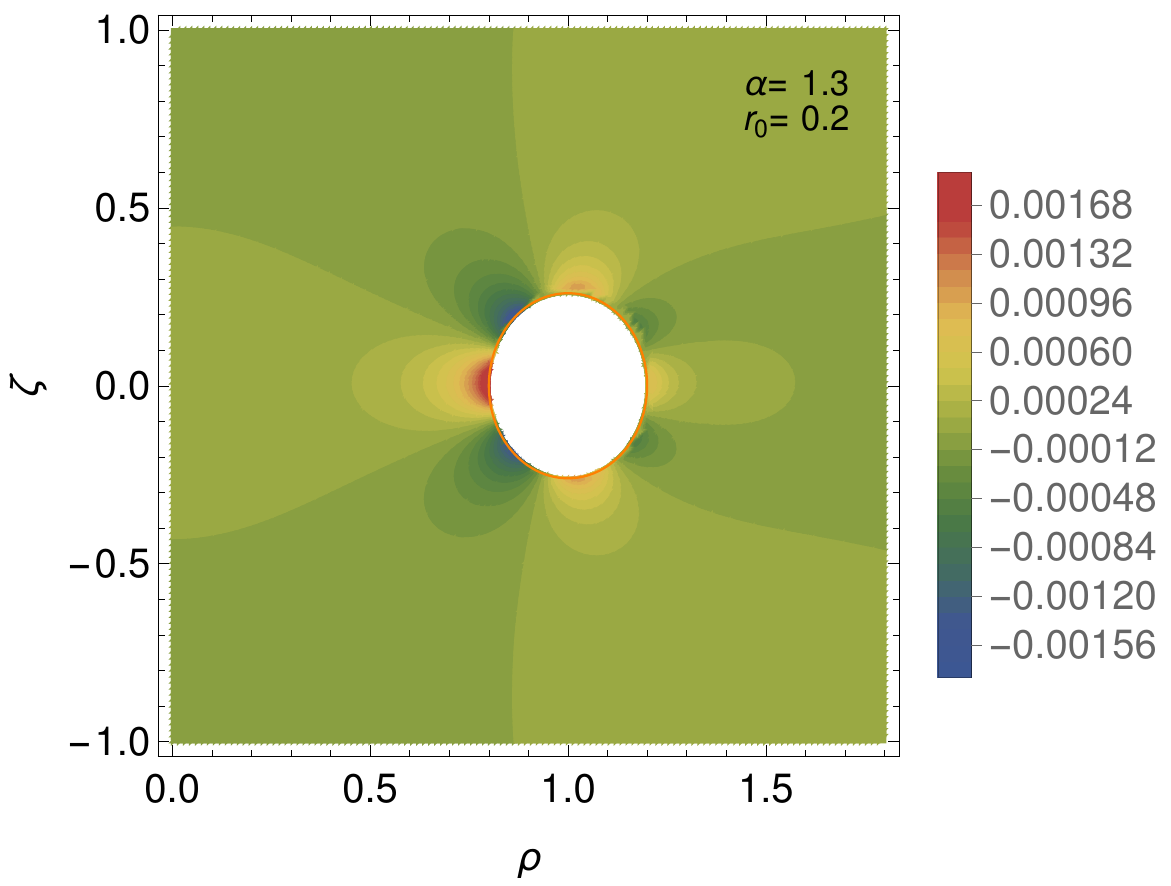}
\includegraphics[width =50mm]{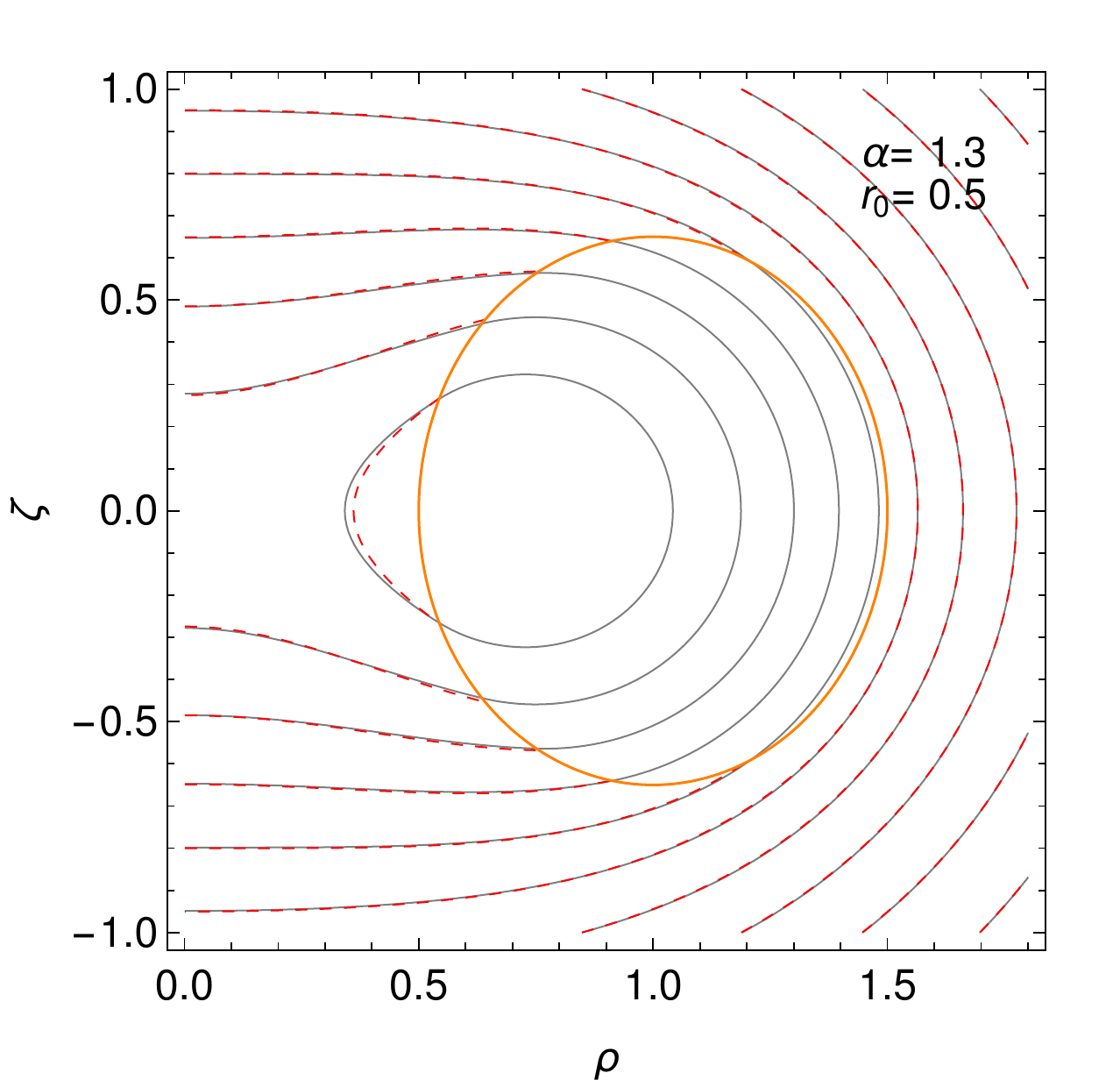}\qquad\qquad 
\includegraphics[width =65mm]{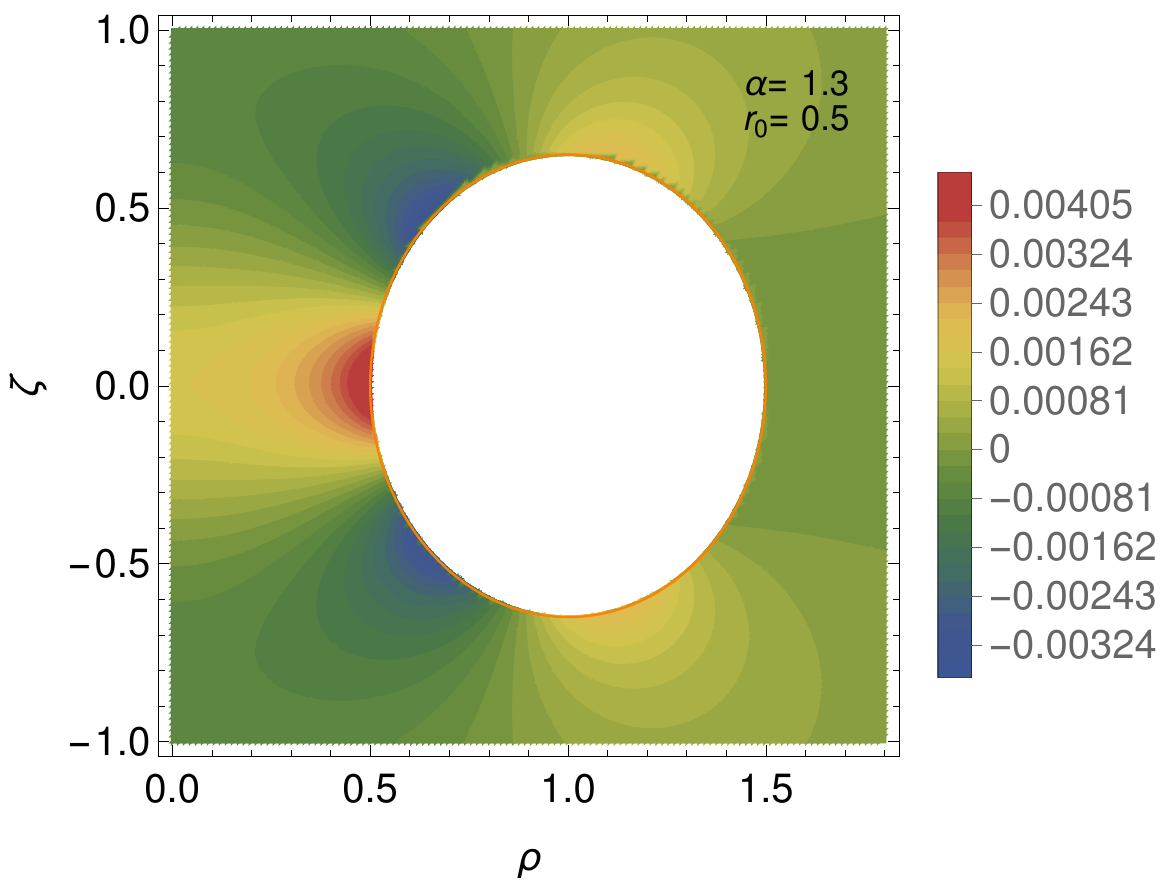}
 \caption{The same that on Fig.\ref{fig:Isoph09} but for $\alpha = 1.3$.}
 \label{fig:Isoph13}
\end{figure*}
The interesting result is that the same formula (\ref{eq4.1}) works also for the case $\alpha > 1$ (prolate cross-section of the torus). In this case, the distances to the massive circles are imaginary
 and conjugate numbers, which looks strange. The explanation of this can be found reminding that complex numbers in Cartesian system are organized in such a way that real numbers correspond to the abscissa axis and imaginary ones to the ordinate. So, when the elliptical cross-section of the torus becomes prolate, the foci are formally on the imaginary axis. As the result, the distances to the massive circle became also imaginary and conjugate (Fig. \ref{fig:ellipse}, {\it Right}) but the resulting values of the potential remain real\footnote{There is some similarity with the outer potential of the spheroid. \cite{1963SvA.....7..276A} considered the problem of motion around two fixed massive points. The masses and distances of these points have the imaginary quantities but the potential is real in any point. They discussed that such a kind of solution can be used to represent the outer potential of the rigid spheroid, for example, the potential of the Earth.}. Let us proof it.  

Indeed, we can represent the distance to massive circles as: 
\begin{equation}\label{eq4.6}
l= i y.
\end{equation}
Substituting (\ref{eq4.6}) in (\ref{eq4.4}), we obtain for the parameters of the elliptical integrals:
\begin{equation}\label{eq4.7a}
  m_{1,2} = \beta \pm i \gamma,  
\end{equation}
where
\begin{equation}\label{eq4.8}
\beta = \frac{4\rho \, (\xi + 2(1+\rho)y^2)}{\xi^2 + 4(1+\rho)^2 y^2},
\end{equation}
\begin{equation}\label{eq4.9}
\gamma = \frac{4\rho \, y\, (\xi - 2(1+\rho))}{\xi^2 + 4(1+\rho)^2 y^2},
\end{equation}
\begin{equation}\label{eq4.9a}
\xi = (1+\rho)^2 + \zeta^2 - y^2.
\end{equation}
As the result, the parameters are conjugate $m_2 = \overline{m}_1$.
The elliptical integrals from conjugate parameters give also the conjugate numbers: 
\begin{align}
  & K(m_1) = \lambda + i \sigma, \label{eq4.10a} \\
  & K(m_2) = \lambda - i \sigma. \label{eq4.10b}
\end{align}
The first multiplier in (\ref{eq4.3}) taking into account (\ref{eq4.7a}) has the form:
\begin{align}\label{eq4.11}
  &\sqrt{\frac{m_1 (1+i y)}{\rho}} = 
  \sqrt{\frac{1}{\rho}\Big[{(\beta -\gamma y) + 
   i (\beta y + \gamma)\Big]}} , \\
   &\sqrt{\frac{m_2 (1 - i y)}{\rho}} = 
  \sqrt{\frac{1}{\rho}\Big[{(\beta -\gamma y) - 
   i (\beta y + \gamma)\Big]}} .
\end{align}
Let us show that they are the complex conjugate quantities and find their expression through coordinates, i.e.:
\begin{align}
  & 
  \sqrt{\frac{1}{\rho}\Big[{(\beta -\gamma y) + 
   i (\beta y + \gamma)\Big]}} = \mu + i \nu , \label{eq4.11aa}\\
   & \sqrt{\frac{1}{\rho}\Big[{(\beta -\gamma y) - 
   i (\beta y + \gamma)\Big]}} = \mu - i \nu . \label{eq4.11bb}
\end{align}
We square the left and the right parts of (\ref{eq4.11aa}) and (\ref{eq4.11bb}). The comparison of the real and imaginary parts gives us the system for both cases: 
\begin{equation}\label{eq4.11dd}
   \begin{dcases}
     \frac{1}{\rho}\Big(\beta - \gamma y\Big)& = \mu^2 - \nu^2 , \\
                \frac{1}{\rho}\Big(\beta y + \gamma\Big) &=  		   2\mu\nu. 
   \end{dcases}
\end{equation}
Expressing $\mu$ from the second equation of (\ref{eq4.11dd}) and putting it in the first one, we obtain:
\begin{equation}\label{eq4.11ee}
\nu^4 + \frac{\beta - \gamma y}{\rho}\nu^2 - \frac{(\beta y + \gamma)^2}{4\rho^2}=0.
\end{equation}
We have only one real and positive root which is:
\begin{equation}\label{eq4.11ff}
\nu = \sqrt{\frac{1}{2\rho}
\Big[ \sqrt{\beta^2 + \gamma^2 + (\gamma y)^2 + (\beta y)^2}+\gamma y - \beta \Big]},
\end{equation}
and
\begin{equation}\label{eq4.11gg}
\mu = \frac{(\beta y+\gamma)}{2\rho\nu}.
\end{equation}
Note, that negative root of (\ref{eq4.11ee})  also satisfies  the conjugation condition (\ref{eq4.11aa}), (\ref{eq4.11bb}) by only changing the sign of $\nu$.
Replacing (\ref{eq4.10a}) -- (\ref{eq4.10b}) and
(\ref{eq4.11aa}) -- (\ref{eq4.11bb})
 in the sum of dimensionless potentials of the two massive circles (the second multiplier in (\ref{eq4.1})) we have:
\begin{align}\label{eq4.12}
 &\phi_{mc,1} + \phi_{mc,2} = (\mu + i \nu)(\lambda + i \sigma) + (\mu - i \nu)(\lambda - i \sigma) = \nonumber \\
 & \qquad \qquad = 2(\mu\lambda - \nu\sigma),
 \end{align}
 where $\mu$ and $\nu$ are functions of the coordinates.
So, the resulting quantity is a real one. Since the last multiplier in (\ref{eq4.1}) is also real, the final value of the potential is real.

 The isopotential curves and corresponding error maps are shown in Figs. \ref{fig:Isoph11} and \ref{fig:Isoph13}. It is apparent that the approximation by two massive circles (\ref{eq4.1}) is working well even for very elongated cross-sections: for the axis ratio $\alpha = 1.6$ and for the large value of $r_0 = 0.7$ the mean absolute error is about 0.3\% and the mean standard deviation about 0.6\%.

\section{Conclusions}
\label{Conclusions}
We present the new integral expression for the homogeneous torus with an elliptical cross-section, working at any point of the space (inside and outside of the torus body). We obtained it by modelling the torus through a set of massive circles, i.e. by the same procedure that we used for the circular cross-section. This result allowed us to find the following new properties: 

\begin{itemize}
\item The outer potential of a homogeneous torus with an elliptical cross-section can be represented with good accuracy by the potential of two massive circles, each of mass equal to half of that of the torus and located along the major axis of torus cross-section in opposite directions with respect to the center. 

\item The distances to these massive circles from the center of elliptical cross-section are half of those to the foci. 

\item
For the prolate case of the torus cross-section, these distances are imaginary and conjugate ones. The imaginary quantities of the massive circle locations lead to the real values of the potential. 

\item The error maps show that the two-circle model for the outer potential of the torus works well for both cases (oblate and prolate cross-sections). 
\end{itemize}
The obtained approximate expression simplifies the treatment of many problem which are related to the investigation of the dynamics in the gravitational field of a torus (the interpretation of N-body simulations for the central region of AGNs, dynamics of stars in the ring galaxies, etc.). 

\section*{Acknowledgements}
\addcontentsline{toc}{section}{Acknowledgements}
We thanks Massimo Capaccioli for the very fruitful discussions which helped us to improve our results and the text of the paper. 
The work was supported under the special program of the NRF of Ukraine "Leading and Young Scientists Research Support" -- "Astrophysical Relativistic Galactic Objects (ARGO): life cycle of active nucleus", No.~2020.02/0346.

\section*{Data availability}
\addcontentsline{toc}{section}{Data availability}
Simulation data and codes used in this paper can be made available upon request by emailing the corresponding author.


\bibliographystyle{mnras}
\bibliography{pottorus} 
\appendix

\section[]{Checking the optimal distances of massive circles}
\label{App}

Here we explain the method that we used to estimate the optimal distances of the two massive circles from the center of the torus cross-section in the representation of the outer torus potential. To do this, we estimate the mean absolute error (MAE) using the obtained relative error maps as
\begin{equation}\label{app1.1}
\text{MAE}(r_0,\alpha, k) = \frac{1}{N^{\prime}}\sum_{i=1}^{N}{\vphantom{\sum}}^\prime\sum_{j=1}^{N}{\vphantom{\sum}}^\prime |\text{RE} (\rho_i, \zeta_j; r_0,\alpha, k)| ,
\end{equation}
where $|\text{RE}|$ is the absolute relative error in each point determined by (\ref{eq3.9b}) for the distance to the massive circles depending on some constant value: \mbox{$l = f/(\alpha + k)$}. The prime attached the summation in (\ref{app1.1}) indicates that we take into account only the $N^\prime$ points outside the torus cross-section fulfilling the condition (\ref{eq4.00}) for the outer region.
 With the same parameters we calculate the mean standard deviation (MSD) of the relative errors as:
 \begin{align}\label{app1.2}
 &\text{MSD}(r_0,\alpha, k)  = \nonumber \\
 & = \sqrt{\frac{1}{N^\prime}\sum_{i=1}^{N}{\vphantom{\sum}}^\prime\sum_{j=1}^{N}{\vphantom{\sum}}^\prime \Big[\text{RE} (\rho_i, \zeta_j;r_0,\alpha, k) - \text{ME}(r_0,\alpha, k)\Big]^2},
 \end{align}
where $\text{ME}$ is the mean error determined by (\ref{app1.1}) but without modulus.
We made a set of the simulations for the some range of $k$ and for each value of $\alpha$ and $r_0$. Some of them are presented in the figures below. It is apparent from Figs \ref{fig:MAE09} to \ref{fig:MAE07} for oblate case and  Figs \ref{fig:MAE11} to \ref{fig:MAE13} for prolate case
that the minimum of the mean absolute error and mean standard deviation for a fixed $\alpha$ and for different values of $r_0$ corresponds always to $k + \alpha = 2$.  

\begin{figure} 
\includegraphics[width =75mm]{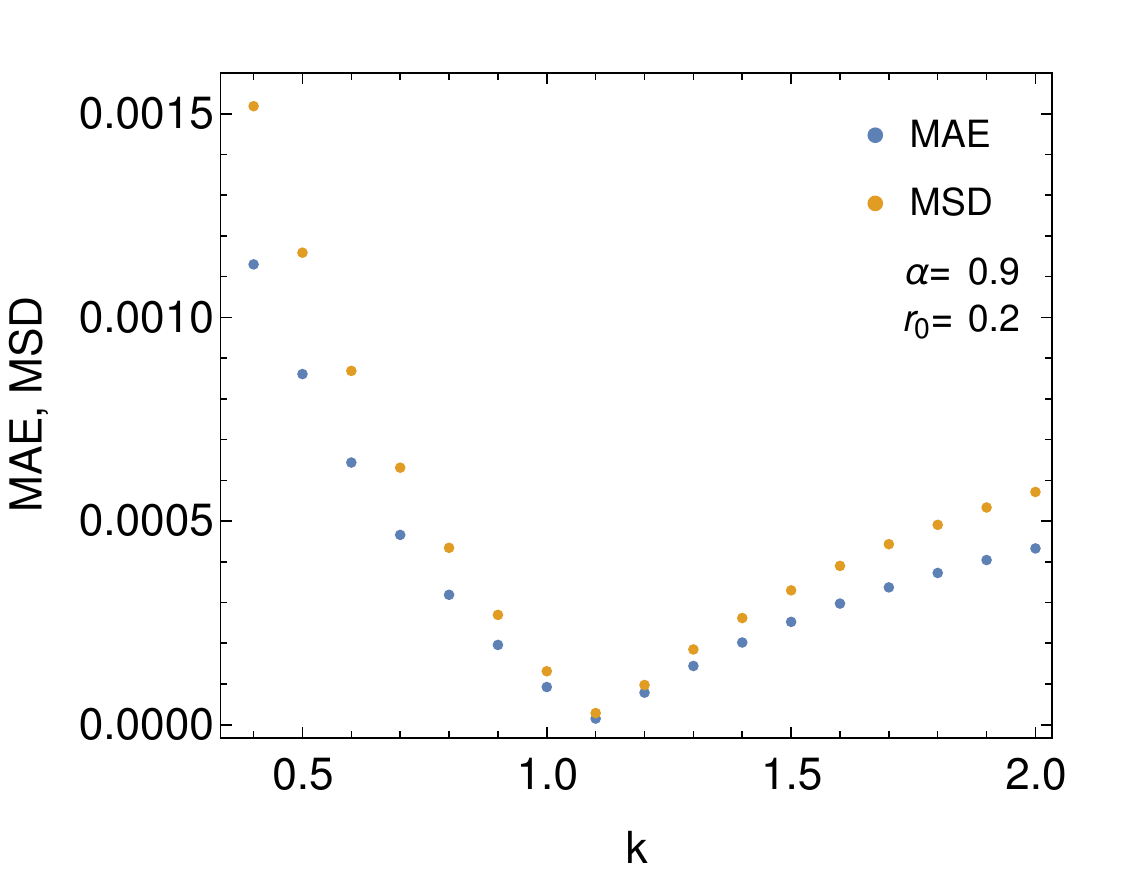}
\includegraphics[width =75mm]{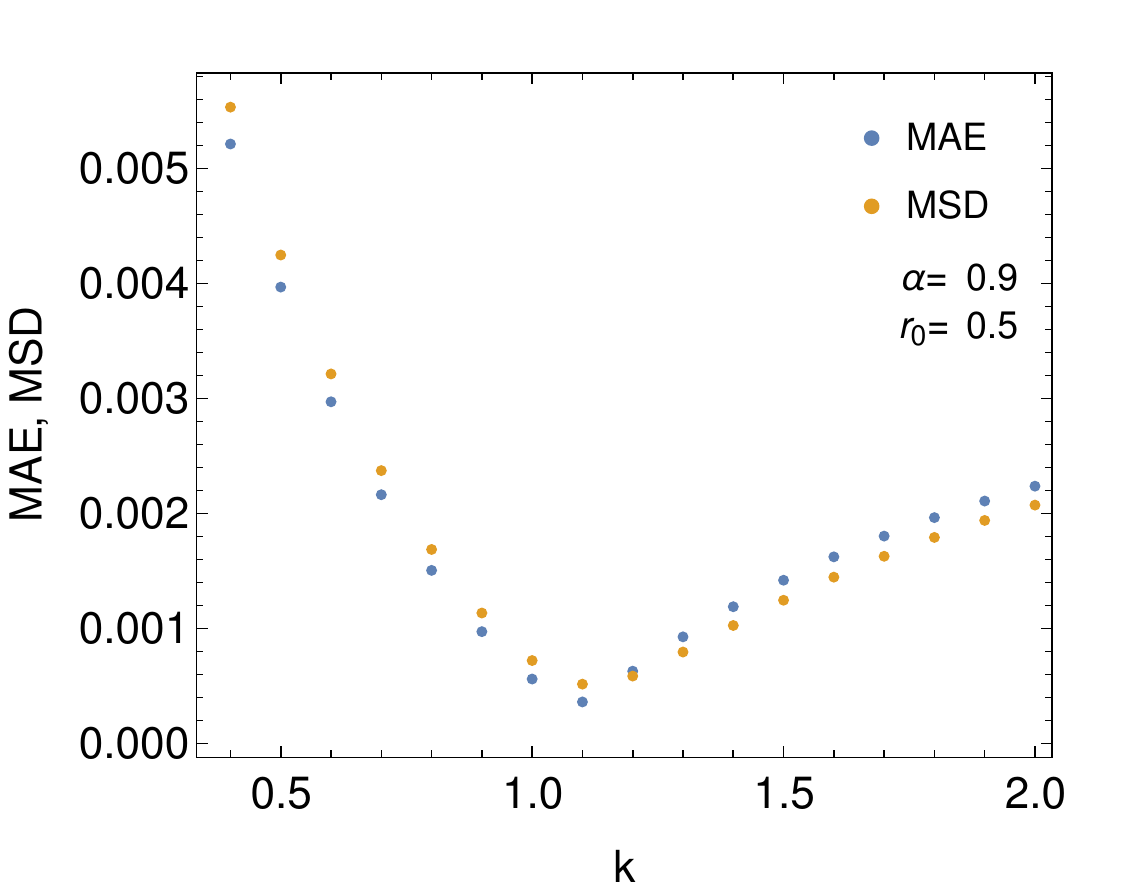}
 \caption{The mean absolute error (MAE) and  the mean standard deviation (MSD) for the parameters: $\alpha = 0.9$ and $r_0 = 0.2$ ({\it top}), $r_0 = 0.5$ ({\it bottom}). The optimal value is $k=1.1$.}
 \label{fig:MAE09}
\end{figure}
\begin{figure} 
\includegraphics[width =75mm]{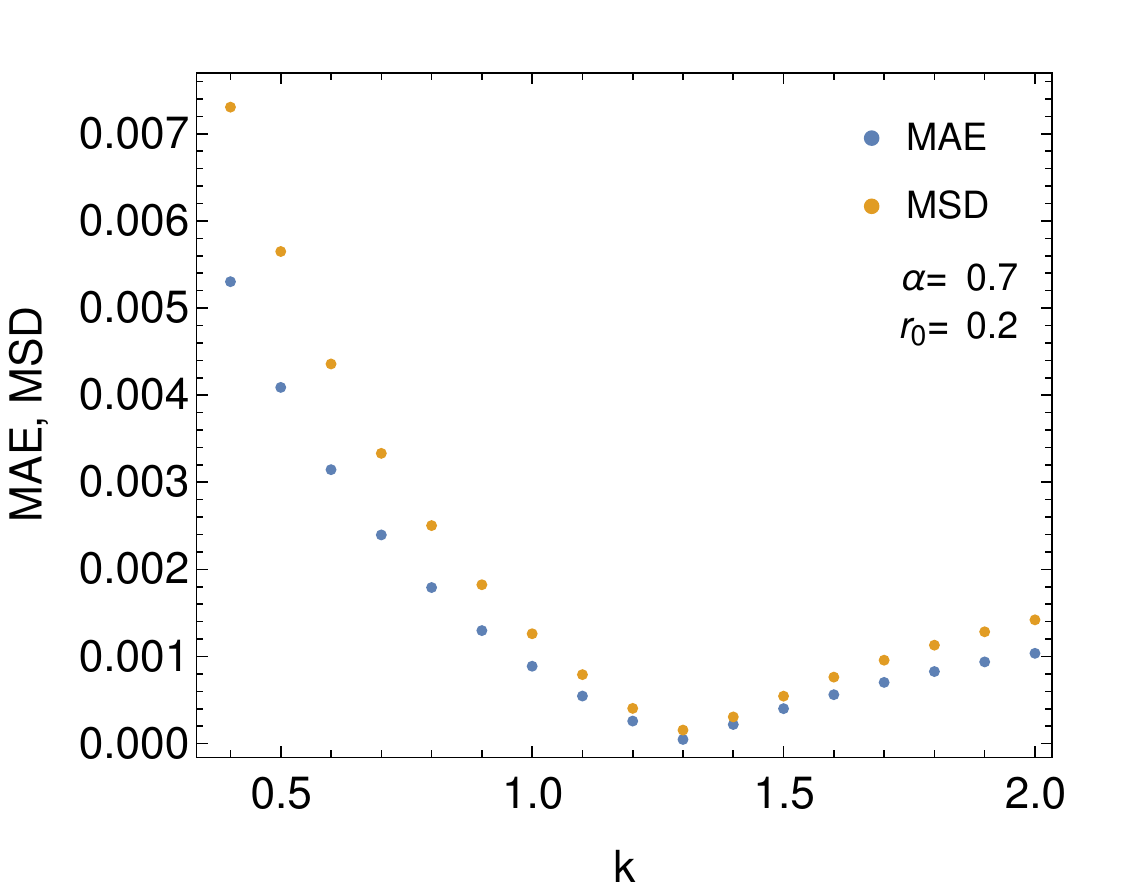}
\includegraphics[width =75mm]{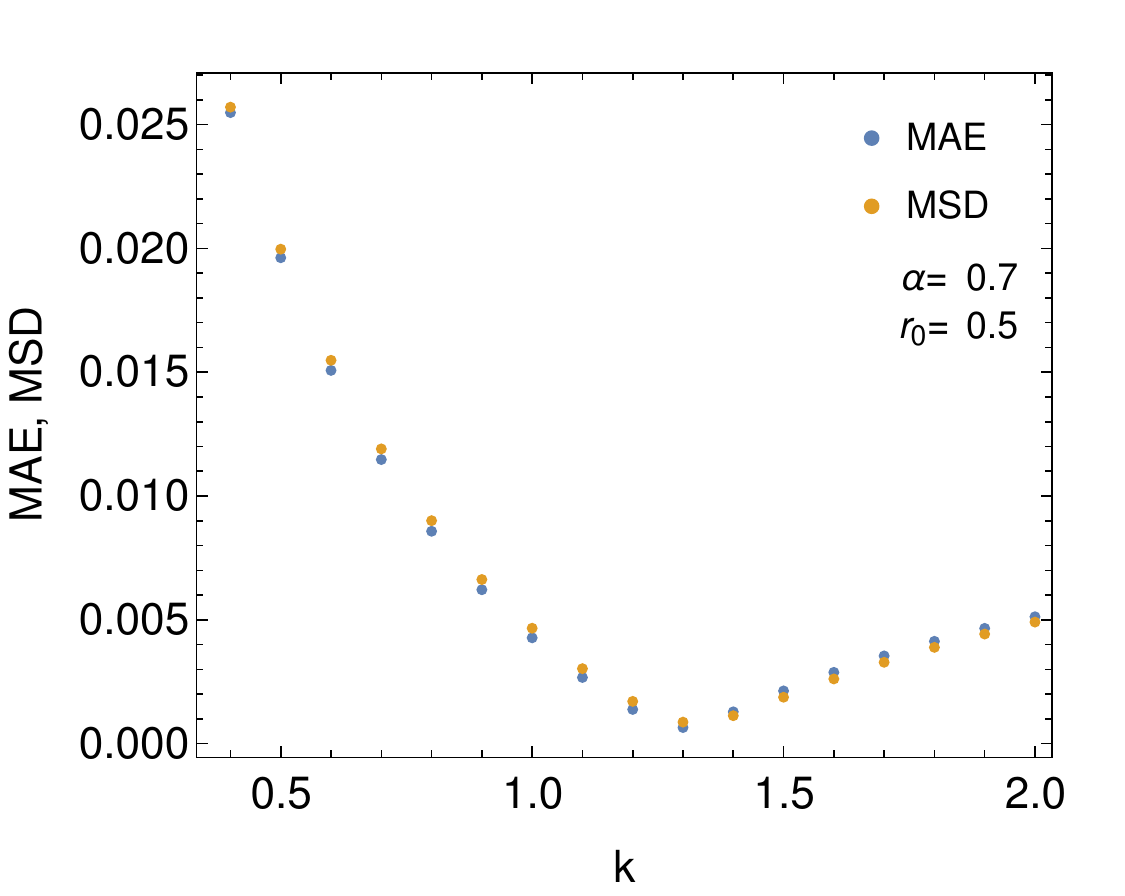}
 \caption{Same as in Fig.~\ref{fig:MAE09} but for $\alpha = 0.7$ and $r_0 = 0.2$ ({\it top}), $r_0 = 0.5$ ({\it bottom}). The optimal value is $k=1.3$.}
 \label{fig:MAE07}
\end{figure}
\begin{figure} 
\includegraphics[width =75mm]{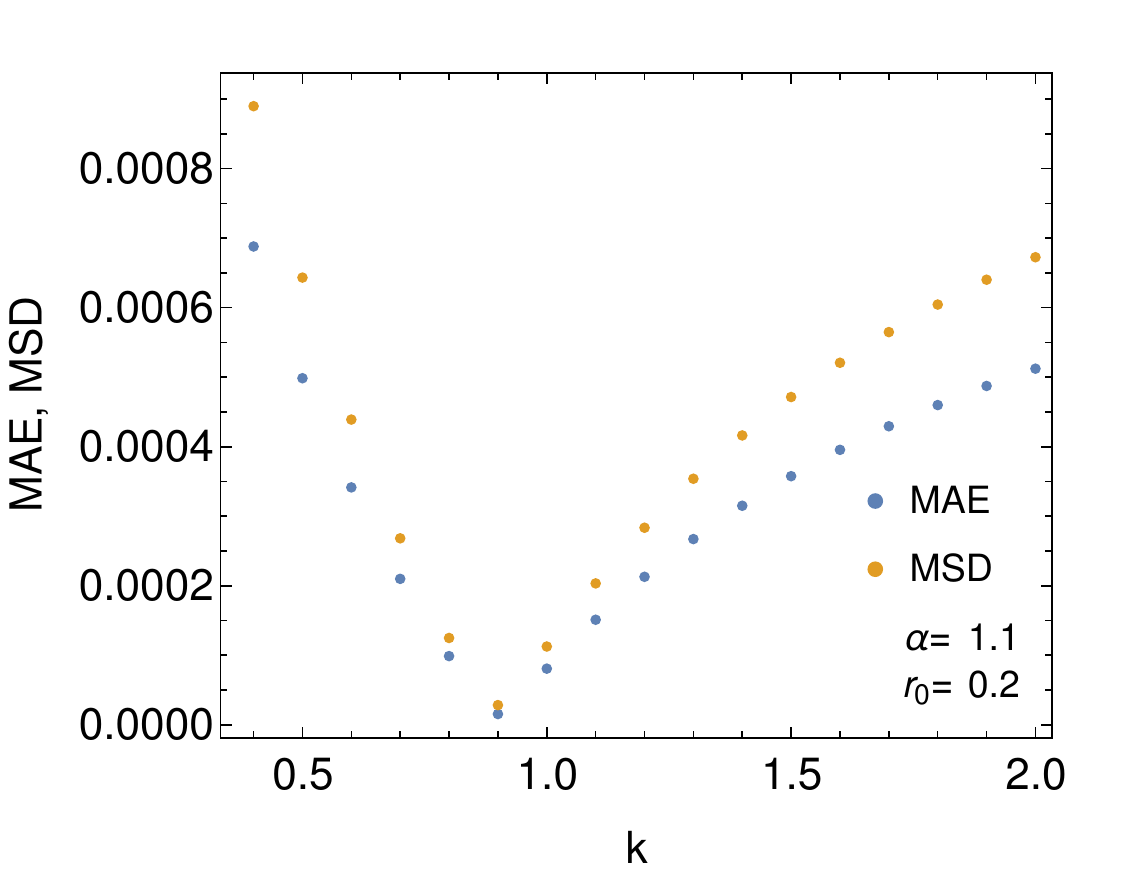}
\includegraphics[width =75mm]{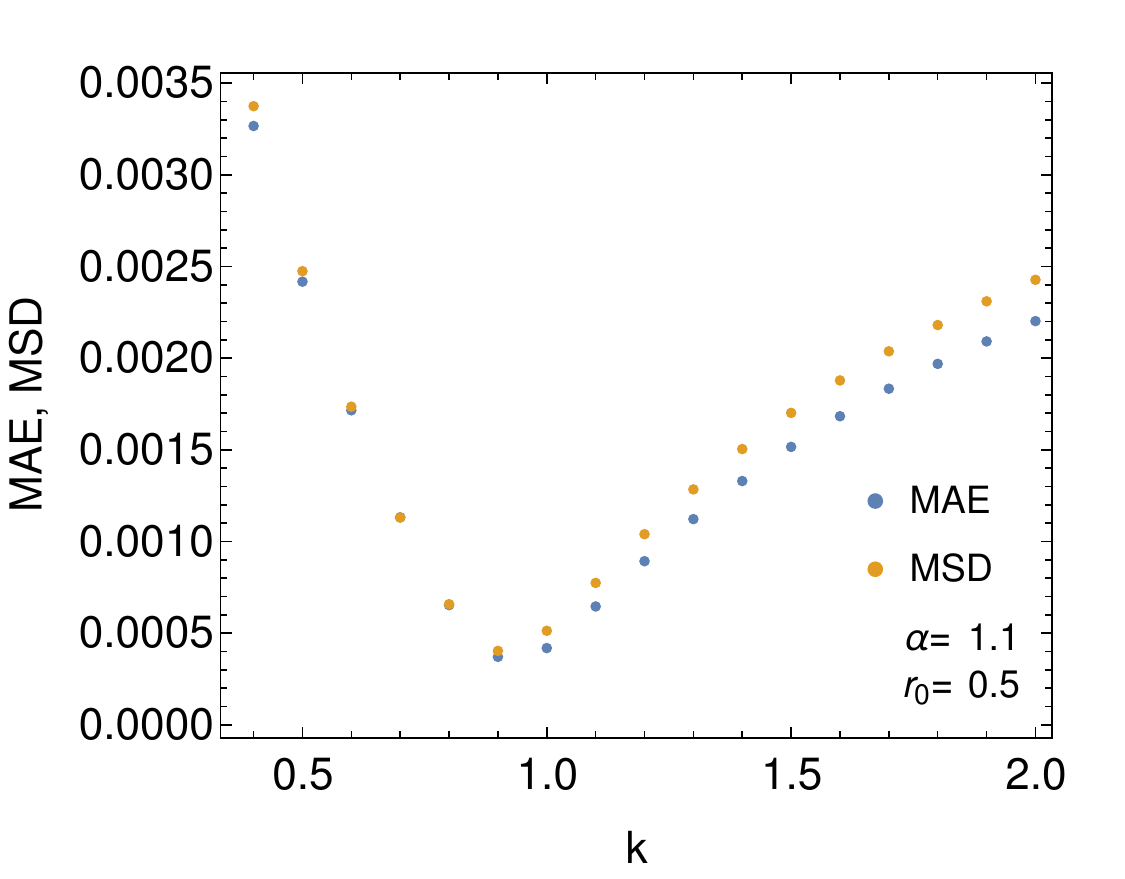}
 \caption{Same as in Fig.~\ref{fig:MAE09} but for $\alpha = 1.1$ and $r_0 = 0.2$ ({\it top}), $r_0 = 0.5$ ({\it bottom}). The optimal value is $k=0.9$.}
 \label{fig:MAE11}
\end{figure}
\begin{figure} 
\includegraphics[width =75mm]{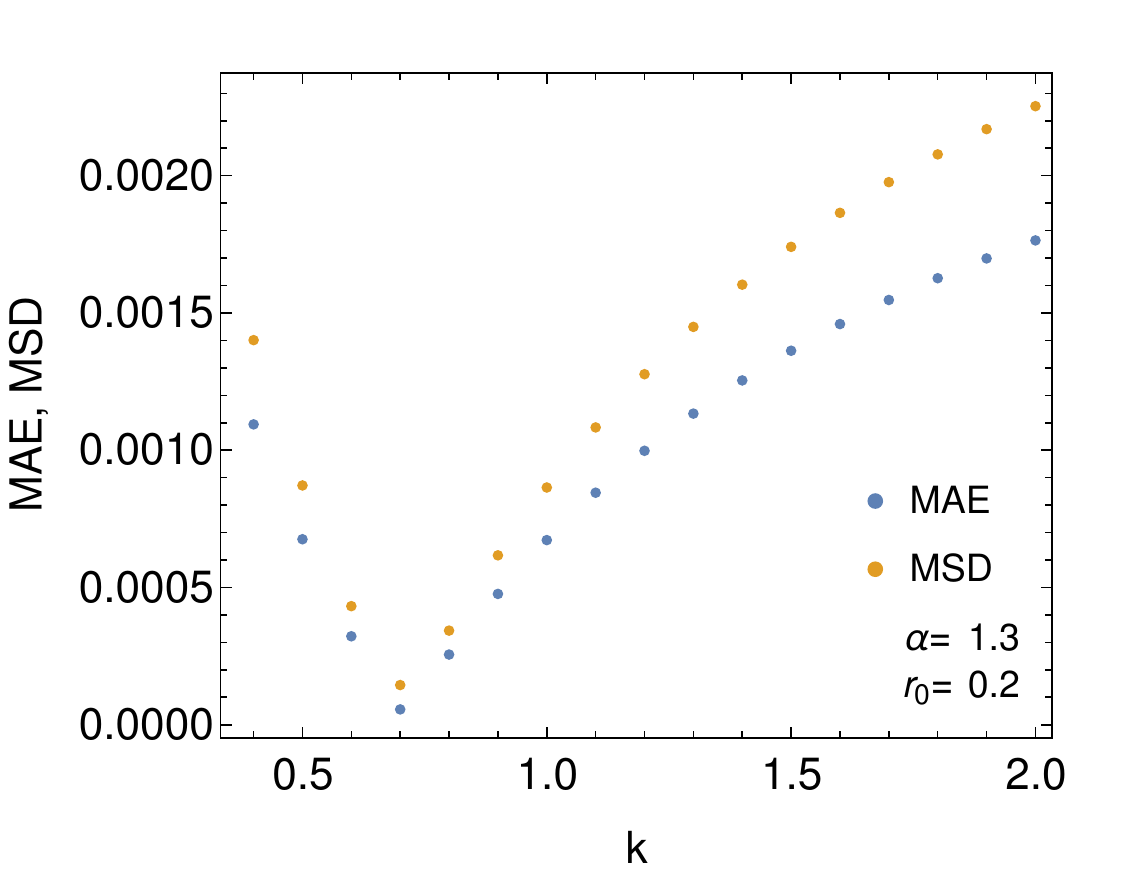}
\includegraphics[width =75mm]{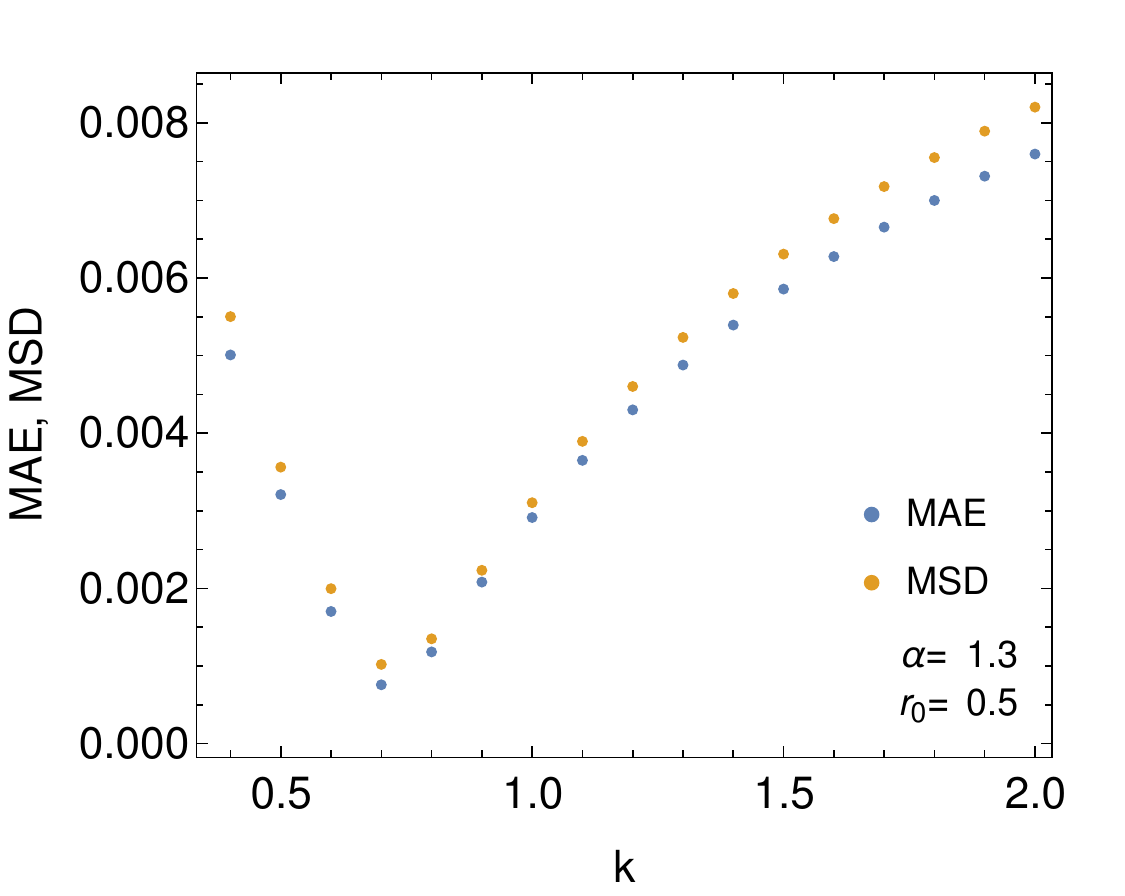}
 \caption{Same as in Fig.~\ref{fig:MAE09} but for  $\alpha = 1.3$ and $r_0 = 0.2$ ({\it top}), $r_0 = 0.5$ ({\it bottom}). The optimal value is $k=0.7$.}
 \label{fig:MAE13}
\end{figure}
\bsp	
\label{lastpage}
\end{document}